\documentclass[submission, Phys]{SciPost}

\usepackage{amsmath,amssymb,mathtools}
\usepackage{color}
\usepackage{graphicx}

\def \be {\begin{equation}}
\def \ee {\end{equation}}
\def \ba {\begin{array}}
\def \ea {\end{array}}
\def \bea{\begin{eqnarray}}
\def \eea{\end{eqnarray}}
\def \nn {\nonumber}

\def \g {\gamma}
\def \G {\Gamma}
\def \d {\delta}

\def \m {\mu}
\def \n {\nu}

\def \l {\lambda}
\def \L {\Lambda}
\def \s {\sigma}

\def \r {\rho}

\def \cO {\mathcal O}

\def \rN {\mathrm N}

\def \f {\frac}

\def \sr {\sqrt}
\def \td {\tilde}

\def \pp {\propto}
\def \inf {\infty}

\def \lag {\langle}
\def \rag {\rangle}

\def \dd {\!\!\mathrm{d}}
\def \ep {\mathrm{e}}
\def \ii {\mathrm{i}}

\def \tr {\textrm{tr}}

\def \diag {{\textrm{diag}}}

\def \and {{~\textrm{and}~}}

\def \CFT {{\textrm{CFT}}}

\def \BW {{\textrm{BW}}}

\begin{document}

\begin{center}{\Large \textbf{
Lattice Bisognano-Wichmann modular Hamiltonian in critical quantum spin chains
}}\end{center}

\begin{center}
Jiaju Zhang\textsuperscript{1,2},
Pasquale Calabrese\textsuperscript{1,2,3},
Marcello Dalmonte\textsuperscript{1,3},
M. A. Rajabpour\textsuperscript{4}
\end{center}

\begin{center}
{\bf 1} Scuola Internazionale Superiore di Studi Avanzati (SISSA),\\Via Bonomea 265, 34136 Trieste, Italy
\\
{\bf 2} INFN Sezione di Trieste, Via Bonomea 265, 34136 Trieste, Italy
\\
{\bf 3} International Centre for Theoretical Physics (ICTP),\\Strada Costiera 11, 34151 Trieste, Italy
\\
{\bf 4} Instituto de Fisica, Universidade Federal Fluminense,\\
Av. Gal. Milton Tavares de Souza s/n, Gragoat\'a, 24210-346, Niter\'oi, RJ, Brazil
\\
\end{center}

\begin{center}
\today
\end{center}


\section*{Abstract}
{\bf
We carry out a comprehensive comparison between the exact modular Hamiltonian and the lattice version of the Bisognano-Wichmann (BW) one in one-dimensional critical quantum spin chains.
As a warm-up, we first illustrate how the trace distance provides a more informative mean of comparison between reduced density matrices
when compared to any other Schatten $n$-distance, normalized or not.
In particular, as noticed in earlier works, it provides a way to bound other correlation functions in a precise manner, i.e., providing both lower and upper bounds.
Additionally, we show that two close reduced density matrices, i.e. with zero trace distance for large sizes, can have very different modular Hamiltonians.
This means that, in terms of describing how two states are close to each other, it is more informative to compare their reduced density matrices rather than the
corresponding modular Hamiltonians.
After setting this framework, we consider the ground states for infinite and periodic XX spin chain and critical Ising chain.
We provide robust numerical evidence that the trace distance between the lattice  BW reduced density matrix
and the exact one goes to zero as $\ell^{-2}$ for large length of the interval $\ell$.
This provides strong constraints on the difference between the corresponding entanglement entropies and correlation functions.
Our results indicate that discretized BW reduced density matrices  reproduce exact entanglement entropies and correlation functions of local operators
in the limit of large subsystem sizes.
Finally, we show that  the BW reduced density matrices fall short of reproducing the exact behavior of
the logarithmic emptiness formation probability in the ground state of the XX spin chain.
}

\vspace{10pt}
\noindent\rule{\textwidth}{1pt}
\tableofcontents\thispagestyle{fancy}
\noindent\rule{\textwidth}{1pt}
\vspace{10pt}

\section{Introduction}

Measures of quantum entanglement and, in particular, entanglement entropies have become one of the key tools to characterize quantum many-body systems
and quantum field theories \cite{Amico:2007ag,Eisert:2008ur,calabrese2009entanglement,Laflorencie:2015eck,Witten:2018lha}.
Given a pure state $|\Psi\rangle$ and a subsystem $A$, the bipartite entanglement is quantified by the von Neumann entropy
\be
S_A = - \tr_A ( \r_A \log \r_A ),
\ee
where the reduced density matrix (RDM) $\r_A=\tr_{\bar A}|\Psi\rangle\langle\Psi|$  is obtained by tracing $|\Psi\rangle\langle\Psi| $ over $\bar A$ the complement of A.
More generally, one can consider the moments of the RDM, i.e.\ $\tr_A \r_A^n$, and define the R\'enyi entropy as
\be
S_A^{(n)} = - \f{\log \tr_A \r_A^n}{n-1} .
\ee
In the $n\to1$ limit, the R\'enyi entropy returns the entanglement entropy; we note that R\'enyi entropies of integer order have already been experimentally measured up to partitions consisting of 10 spins
~\cite{daley2012measuring,islam2015measuring,elben2018renyi,vermersch2018unitary,brydges2019probing,tta}.
 The RDM $\r_A$ is fully encoded in the modular (or entanglement) Hamiltonian $H_A$ defined as
\be
\r_A = \f{\ep^{- H_A}}{Z_A}, \qquad Z_A = \tr_A \ep^{- H_A}.
\ee
By construction, the RDM and the modular Hamiltonian have the same eigenvectors, and their eigenvalues are simply related. The modular Hamiltonian plays a key role in quantum field theory~\cite{Witten:2018lha}, and recently attracted a great amount of interest in the context of condensed matter physics, in particular, for
topological matter \cite{li2008entanglement}.

While, in the context of relativistic quantum field theory, the modular Hamiltonian of half-space partition is known to be related to the boost operator~\cite{Bisognano:1975ih,Bisognano:1976za}, its explicit functional form in lattice models is hard to construct, and is known only in a few simple 
cases
\cite{itoyama1987lattice,peschel1999density,nienhuis2009entanglement,peschel2009reduced,eisler2017analytical,eisler2018properties,Eisler:2019rnr,DiGiulio:2019cxv}.
In order to overcome this challenge, it was proposed in Refs.~\cite{Dalmonte:2017bzm,Giudici:2018izb} to use the Bisognano-Wichmann (BW) theorem
in quantum field theory \cite{Bisognano:1975ih,Bisognano:1976za} and its extension in conformal field theory (CFT)
\cite{Hislop:1981uh,Casini:2011kv,Wong:2013gua,Wen:2016inm,Cardy:2016fqc,Najafi:2016kwb,Fries:2019acy,Fries:2019ozf,Tonni:2017jom,DiGiulio:2019lpb} to write approximate modular Hamiltonians for lattice models.
From the BW modular Hamiltonian one can construct a RDM, which has been dubbed BW RDM.
The proposal has been checked extensively
\cite{Kim:2016cdh,Dalmonte:2017bzm,Giudici:2018izb,ParisenToldin:2018uzz,Kosior:2018vgx,eisler2018properties,Zhu:2019tsb,Turkeshi:2018hfx,Mendes-Santos:2019ine,Mendes-Santos:2019tmf},
showing that in many cases the BW modular Hamiltonian can reproduce to a good precision the entanglement spectrum, correlation functions, entanglement entropy and R\'enyi entropies.

In this paper we further investigate the precision of the lattice BW modular Hamiltonian when {\it the size of the subsystem becomes large}.
It was found in \cite{eisler2017analytical} that for a free fermion chain the deviation of the BW modular Hamiltonian
from the exact one persists even for a large subsystem.
However, two RDMs with asymptotically zero distance can have very different modular Hamiltonians,
as we will show in section~\ref{sectoy} using a toy example, and so the modular Hamiltonian itself
may not be a good quantity to distinguish different states of the subsystem.
Recently in \cite{Mendes-Santos:2019tmf}, the Schatten distances, with a normalization
as in \cite{fagotti2013reduced}, between the BW RDM and the exact RDM were calculated, and it
was found that they decay algebraically with interval's size.
In this paper we want to study the same problem using the trace distance.
There are at least two good reasons to study also the trace distance between the two reduced density matrices.
First, the behaviour of trace distance between the BW and the exact RDMs imposes strong constraints on the behaviour of other quantities such as correlation functions, entanglement entropy, and R\'enyi entropies through various inequalities.
The verification of these operator inequalities was so far essentially neglected.
Second, as we will show in section~\ref{sectoy}, there exist rather different states that the Schatten distance
of a large subsystem, normalized or not, cannot distinguish, while the trace distance can.

Since in this article we just use the BW modular Hamiltonian of two-dimensional (2D) CFTs,
we will focus on two critical points of the XY spin chain, i.e., the XX spin chain with zero magnetic field and the critical Ising spin chain.
The XY chain can be exactly diagonalized and this allows us to calculate the trace distance
between the BW and the exact RDMs for one interval with length as large as $\ell\sim100$.
Furthermore, we also calculate the fidelity, which can give an upper
bound of the trace distance, for intervals up to size $\ell\sim1000$ for XX chain and $\ell\sim500$ for critical Ising chain.
We find a power-law decay of the trace distance in $\ell$.
Then, exploiting the inequalities satisfied by the trace distance, we will look into the behavior of entanglement entropy,
R\'enyi entropy, RDM moments, correlation functions, and formation probabilities.


The remaining part of the paper is arranged as follows. In section~\ref{secTraceDis}
we first define the trace distance and then summarize its relevant properties.
In section~\ref{sectoy}, we use toy examples to show that the trace distance can distinguish states
that any other $n$-distance, normalized or not, cannot; we also show that two close RDMs in general can have  different modular Hamiltonians.
In section~\ref{secXX} and section~\ref{secIsing}, we consider
the ground state of the XX spin chain with a zero transverse field and of the critical Ising spin chain on an infinite straight line and on a circle.
We conclude with discussions in section~\ref{seccon}.


\section{Trace distance and its properties}\label{secTraceDis}

For two density matrices $\r,\s$ with $\tr\r=\tr\s=1$, the Schatten $n$-distance with $n\geq1$ is defined as \cite{nielsen2010quantum,watrous2018theory}
\be
D_n(\r,\s) = \f{1}{2^{1/n}} \| \r - \s \|_n,
\ee
where for a general matrix $\L$, Schatten $n$-norm with $n\geq1$ is
\be
\|\L\|_n = \Big( \sum_i \l_i^n \Big)^{1/n},
\ee
where $\l_i$'s are the singular values of $\L$, i.e.\ the nonvanishing eigenvalues of $\sr{\L^\dag \L}$. When $\L$ is Hermitian, $\l_i$'s
are the absolute values of the nonvanishing eigenvalues of $\L$.
We have chosen the normalization such that for two orthogonal pure states we have $D_n=1$.
The $n$-distance is not always able to distinguish different states, and in \cite{fagotti2013reduced} an alternative definition of the $n$-distance has been proposed%
\footnote{The $n=2$ version of the normalized $n$-distance (\ref{Dpnrs}) was proposed and used in \cite{fagotti2013reduced} to
investigate the time evolution of
the RDM after a global quench \cite{Calabrese:2005in,Calabrese:2006rx}.
As stated in \cite{fagotti2013reduced}, the triangle inequality for the normalized $n$-distance has not been proven yet, and so it
may even not be a real well-defined distance.}
\be \label{Dpnrs}
\td D_n(\r,\s) = \f{\| \r-\s \|_n}{( \tr\r^n + \tr\s^n )^{1/n}},
\ee
which can be called normalized $n$-distance.
For $n=1$, the $n$-distance becomes the trace distance
\be
D(\r,\s) = \f{1}{2} \| \r - \s \|_1.
\ee
As we will show in section~\ref{sectoy}, the trace distance can distinguish, in a way that we specify in detail below, some states of a large subsystem that any
other $n$-distance, normalized or not, cannot. This is in agreement with the known metric properties of the trace distance
(see, e.g., Ref.~\cite{Gilchrist2005distance, Liang:2018yey}); it also confirms some observations done for excited states of CFTs \cite{Zhang:2019wqo,Zhang:2019itb}, as well as out of equilibrium \cite{Zhang:2019kwu}.

The fidelity of two density matrices $\r$ and $\s$ are defined as \cite{nielsen2010quantum,watrous2018theory}\footnote{In quantum information literature, Eq. (\ref{fidelitydef}) is sometimes called square root fidelity and the fidelity is defined as $F(\r,\s) = \big( \tr \sr{\sr{\s}\r\sr{\s}} \big)^2$.}
\be \label{fidelitydef}
F(\r,\s) = \tr \sr{\sr{\s}\r\sr{\s}}.
\ee
Although not obvious by definition, the fidelity is symmetric to its inputs.
The fidelity provides both a lower bound and an upper bound on the trace distance
\be \label{fidelity-tracedistance}
1 - F(\r,\s) \leq D(\r,\s) \leq \sr{1 - F(\r,\s)^2}.
\ee
The upper bound will be extremely useful to us.


The trace distance also bounds other interesting quantities
such as the entanglement entropy, R\'enyi entropies, RDM moments, and the correlation functions.
The interested reader can find all necessary details in the review \cite{chehade2019quantum}.
The Fannes-Audenaert inequality provides an upper bound for the entanglement entropy difference \cite{Fannes1973,Audenaert:2006}
\be \label{FAI}
|S_A(\r_A) - S_A(\s_A)| \leq D \log (d_A-1) -D\log D -(1-D)\log(1-D),
\ee
where $D \equiv D(\r_{A},\s_{A})$ is the trace distance and $d_A$ is the dimension of the RDM.
For one interval with $\ell$ sites (e.g. in the XY spin chain), one has $d_A=2^\ell$. Hence, if the trace distance decays faster than ${1}/{\ell}$,
then difference of the von Neumann entropies goes to zero for the large subsystem sizes.

The trace distance also puts bounds on the difference of the R\'enyi entropies and RDM moments.
For R\'enyi entropies with $0<n<1$ one has \cite{Audenaert:2006}
\be \label{ieqSn1}
| S_A^{(n)}(\r_A) - S_A^{(n)}(\s_A) | \leq \f{1}{1-n} \log [ (1-D)^n + (d_A-1)^{1-n} D^n ],
\ee
while for $n>1$ \cite{chen2017sharp}
\be \label{ieqSn2}
| S_A^{(n)}(\r_A) - S_A^{(n)}(\s_A) | \leq \f{d_A^{n-1}}{n-1} \Big[ 1 - (1-D)^n - \f{D^n}{(d_A-1)^{n-1}} \Big].
\ee
According to both inequalities, the trace distance should decay exponentially fast in $\ell$ to get a vanishing upper bound.
It is worth mentioning that both inequalities are known to be sharp \cite{chen2017sharp}.
We note that since in 2D CFTs we have    \cite{Holzhey:1994we,Calabrese:2004eu}
\be \label{EEcft}
S^{(n)}_{A,\CFT}(\ell)=\f{c}{6}\Big(1+\f{1}{n}\Big)\log \ell +\gamma_n
\ee
for critical systems the quantity $\big|1-{S_A^{(n)}(\ell)}/{S^{(n)}_{A,\CFT}(\ell)}\big|$ can be bounded by zero if: a) for $n=1$ the trace distance goes to zero like $\frac{1}{\ell^{\alpha}}$ with $\alpha\geq1$; b) for $n\neq1$ the trace distance goes to zero exponentially fast.

For RDM moments, there are also bounds. For $0<n<1$ one has \cite{Audenaert:2006}
\be \label{ieqTn1}
|\tr_A \r_A^n - \tr_A \s_A^n | \leq   (1-D)^n + (d_A-1)^{1-n} D^n - 1,
\ee
and for $n>1$ \cite{raggio1995properties}
\be \label{ieqTn2}
| \tr_A \r_A^n - \tr_A \s_A^n | \leq \f{2D}{n}.
\ee
The last equation implies that, when the trace distance decays to zero for the large subsystems, the difference of the RDM moments also goes to zero for $n>1$. The same conclusion is not true for $0<n<1$.
 Anyhow, in order to have a meaningful constraint, we should also take into account how $\tr_A \r_A^n$ itself scales with $\ell$. For 2D CFTs
 the moment scales as $\tr_A\r_{A,\CFT}^n \pp \ell^{-\f{c}{6}(n-\f{1}{n})}$ which means the quantity $\big|1-\f{\tr_A\r_A^{n}}{\tr_A\r_{A,\CFT}^{n}}\big|$ can be bounded by zero if: a) for $n<1$ the trace distance goes to zero exponentially fast; b) for $n>1$ the trace distance scales like $\ell^{-\alpha}$ with $\alpha>\f{c}{6}(n-\f{1}{n})$.

Finally, the trace distance gives the following constraint on the difference of the expectation value of an operator\cite{nielsen2010quantum}
\be \label{inequality}
| \tr_A [ (\r_A-\s_A)\cO ] | \leq s_{\rm max}(\cO) \| \r_A - \s_A \|_1,
\ee
where $s_{\rm max}(\cO)$ is the largest singular value of $\cO$.
Hence, if the trace distance between two density matrices goes to zero for large subsystem sizes, then the difference between the expectation values of the operators, with finite largest singular value, calculated using the two density matrices goes to zero.
Since most of the local operators in quantum spin chains have finite $s_{\rm max}(\cO)$, the trace distance puts a strong constraint on their value.

\section{Different ways of comparing density matrices: Toy examples}\label{sectoy}

In this section, we give three toy examples that illustrate the comparative predictive power of different distances between density matrices.
The goals of this section are to (1) emphasize the difference between trace and Schatten distances, and (2) point out how states which are very close under trace distance may be described by very distinct modular Hamiltonians.

\paragraph{First example. -} In this first toy example, we show that the trace distance can distinguish two states of a large subsystem that any other $n$-distance cannot.
Since the Schatten norm has the monotonicity property, this example may not be surprising at all; however, it will help us to build up the basis for
the next two toy examples and for the main arguments. We consider the two $2^\ell \times 2^\ell$ diagonal RDMs:
\be \label{egrAsA}
\r_A = \diag( 2^{-\ell}, \cdots, 2^{-\ell} ), \qquad
\s_A = \diag( 2^{-\ell+1}, \cdots, 2^{-\ell+1}, 0, \cdots,0  ).
\ee
The RDM $\r_A$ has $2^\ell$ identical eigenvalues equal to $2^{-\ell}$.
The RDM $\s_A$ has half of its eigenvalues (i.e. $2^{\ell-1}$) equal to $2^{-\ell+1}$ and the other half are vanishing.
Both RDMs are normalized, $\tr_A\r_A=\tr_A\s_A=1$.
The two matrices  are very different, and we expect a finite distance between them.
It is easy to imagine physical observables being very different in the two cases, for example in a spin system.
The trace distance is
\be
D(\r_A,\s_A) = \f12,
\ee
and the normalized $n$-distance (\ref{Dpnrs}) is
\be
\td D_n(\r_A,\s_A) = \f{1}{(1+2^{n-1})^{1/n}},
\ee
both giving a finite value in the large $\ell$ limit. Conversely, the un-normalized $n$-distance is
\be
D_n(\r_A,\s_A) = \f{1}{2^{(1-\f1n)\ell+\f1n}}.
\ee
For $n>1$, $D_n(\r_A,\s_A)$ is exponentially small as $\ell\to\inf$, although the two states are different.
We conclude that the trace distance and the normalized distances (\ref{Dpnrs}) distinguish these two states for a large subsystem, while the $n$-distance cannot.

\paragraph{Second example. -} In this second example, we show that the trace distance can distinguish two states of a large subsystem that other normalized $n$-distances cannot.
Specifically, we consider:
\be
\r_A' = \f12 ( \r_{A,1} + \r_{A} ),  \qquad
\s_A' = \f12 ( \r_{A,1} + \s_{A} ),
\ee
where $\r_{A,1} = \diag(1,0,\cdots,0)$ and $\r_{A},\s_{A}$ are defined in (\ref{egrAsA}).
We get a finite trace distance
\be
D( \r_A', \s_A' ) = \f14.
\ee
Conversely, for $n>1$, the normalized $n$-distance  decays exponentially for large $\ell$ as
\be
\td D_n( \r_A', \s_A' ) \approx \f{1}{2^{(1-\f1n)\ell+\f1n}}.
\ee
We conclude that the trace distance distinguishes two states of a large subsystem that any other normalized $n$-distance can not.

\paragraph{Third example.-} In this last example, we show that two very close RDMs can have different modular Hamiltonians.
We consider:
\bea
&& \r_A = \diag( 2^{-\ell+1}-2^{-\ell^2}, \cdots, 2^{-\ell+1}-2^{-\ell^2}, 2^{-\ell^2}, \cdots , 2^{-\ell^2}), \nn\\
&& \s_A = \diag( 2^{-\ell+1}-2^{-2\ell^2}, \cdots, 2^{-\ell+1}-2^{-2\ell^2},  2^{-2\ell^2}, \cdots, 2^{-2\ell^2} ).
\eea
Half of the eigenvalues of $\r_A$ are $2^{-\ell+1}-2^{-\ell^2}$ and the other half are $2^{-\ell^2}$.
Half of the eigenvalues of $\s_A$ are $2^{-\ell+1}-2^{-2\ell^2}$ and the other half are $2^{-2\ell^2}$.
These two RDMs are very close in the $\ell \to \inf$ limit, and indeed we have an exponentially small trace distance
\be
D(\r_A,\s_A) \approx 2^{-\ell^2+\ell-1} \to 0 \quad {\rm as} \; \ell \to \inf.
\ee
However, they lead to very different modular Hamiltonians
\bea
&& - \log \r_A \approx \diag( (\ell-1)\log2 ,\cdots, (\ell-1)\log2, \ell^2\log2, \cdots, \ell^2\log2), \nn\\
&& - \log \s_A \approx \diag( (\ell-1)\log2 , \cdots, (\ell-1)\log2, 2\ell^2\log2, \cdots, 2\ell^2\log2).
\eea
We conclude that, while a direct comparison of modular Hamiltonians and element-by-element entanglement spectra is certainly informative, the trace distance provides a more informative mean of comparing density matrices for large subsystems.

\section{The XX spin-chain}\label{secXX}

The XX spin chain (in zero field) is described by the Hamiltonian
\be
H_{\rm XX} = - \f14 \sum_{j=1}^L \big( \s_j^x\s_{j+1}^x + \s_j^y\s_{j+1}^y \big).
\ee
Its continuum limit is a free massless compact boson theory, with the target space being a unit radius circle, which is a 2D CFT with central charge $c=1$.
The Hamiltonian of the XX spin chain is mapped to that of free fermions by the Jordan-Wigner transformation
\be \label{JWtransf}
a_j = \Big(\prod_{i=1}^{j-1}\s_j^z\Big)\s_j^+, \qquad
a_j^\dag = \Big(\prod_{i=1}^{j-1}\s_j^z\Big)\s_j^-,
\ee
with $\s_j^\pm = \f12 ( \s_j^x \pm \ii \s_j^y )$.
One can also define the Majorana operators:
\be \label{MajoranaModes}
d_{2j-1}=a_j + a^\dag_j, ~~
d_{2j}=\ii(a_j - a^\dag_j).
\ee
The exact modular Hamiltonian of an interval $A$ of  length $\ell$ takes the form
\be
H_A = \sum_{j_1,j_2=1}^\ell H_{j_1j_2}a_{j_1}^\dag a_{j_2}.
\ee
The RDM can be constructed from the two-point correlation function matrix \cite{peschel2009reduced,chung2001density,cheong2004many,Vidal:2002rm,peschel2003calculation,Latorre:2003kg,peschel2012special} and
$H_A$ is related to $C$ as \cite{cheong2004many,peschel2003calculation}
\be
C = \f{1}{1+\ep^{H_A}}.
\ee
In the XX spin chain, the $\ell\times\ell$ correlation matrix $C$  has entries
\be \label{Cdef}
C_{j_1j_2} = \lag a^\dag_{j_1} a_{j_2} \rag = f_{j_2-j_1}.
\ee
The function $f_j$ takes different forms for different states and geometries.
We only consider ground states.
For a finite interval $A$ in an infinite chain we have
\be \label{fjline}
f_j^{\infty} = \f{1}{\pi j}\sin\f{\pi j}{2}, \qquad  f_0^{\infty}=\f12,
\ee
while for a periodic chain of length $L$ (with $L$ even)
\be \label{fjcircle}
f_j^{\rm PBC} = \f{\sin\f{\pi j}{2}}{L\sin\f{\pi j}{L}}, \qquad f_0^{\rm PBC}=\f12.
\ee

Approximate RDM on the lattice \cite{Dalmonte:2017bzm,Giudici:2018izb} can be constructed following the
BW theorem \cite{Bisognano:1975ih,Bisognano:1976za} and its extensions for CFT
\cite{Hislop:1981uh,Casini:2011kv,Wong:2013gua,Wen:2016inm,Cardy:2016fqc,Najafi:2016kwb,Fries:2019acy,Fries:2019ozf,Tonni:2017jom,DiGiulio:2019lpb}.
For a recent review of BW theorem in quantum field theory, see Ref.~\cite{Witten:2018lha}. For the ground state of an arbitrary $(d+1)$-dimensional relativistic quantum field theory,
the BW theorem states that the modular Hamiltonian of the half-infinite space $A$ (defined by the condition $A=[0,\infty]$) can be
written as the partial Lorentz boost generator
\be
H_A = 2\pi \int_{x\in A} \dd^d x \;x_1 H(x),
\ee
where $H(x)$ is the Hamiltonian density of the theory, and the speed of light has been set to unit.
For a 2D CFT, the BW theorem can be extended to other geometries
\cite{Hislop:1981uh,Casini:2011kv,Wong:2013gua,Wen:2016inm,Cardy:2016fqc,Najafi:2016kwb,Fries:2019acy,Fries:2019ozf,Tonni:2017jom,DiGiulio:2019lpb}.
For a finite interval $A=[0,\ell]$ on an infinite line in the ground state, the modular Hamiltonian is \cite{Casini:2011kv,Tonni:2017jom}
\be
H_A = 2\pi \int_0^\ell \dd x \f{x(\ell-x)}{\ell} H(x).
\ee
For the interval $A = [0,\ell]$ in the ground state of a periodic system of total length  $L$, the modular Hamiltonian is \cite{Tonni:2017jom}
\be
H_A = 2\pi \int_0^\ell \dd x \f{\sin\f{\pi x}{L}\sin\f{\pi (\ell-x)}{L}}{\f{\pi}{L}\sin\f{\pi\ell}{L}} H(x).
\ee
We now briefly review how to adapt the continuum formulation to ground states of lattice models~\cite{Dalmonte:2017bzm,Giudici:2018izb}. The BW
RDM of a given interval is
\be
\r_A^\BW = \f{\ep^{- H_A^\BW}}{Z_A^\BW}, \qquad Z_A^\BW = \tr_A \ep^{- H_A^\BW}.
\ee
One can use the modular Hamiltonian in 2D CFT \cite{Hislop:1981uh,Casini:2011kv,Wong:2013gua,Wen:2016inm,Cardy:2016fqc} to write the BW modular Hamiltonian as
\be \label{HABWdef}
H_A^\BW = \sum_{j_1,j_2=1}^\ell H_{j_1j_2}^\BW a_{j_1}^\dag a_{j_2},
\ee
where matrix $H^\BW$ has only nearest neighbor non-vanishing entries
\be \label{HBWjjp1}
H^\BW_{j,j+1} = H^\BW_{j+1,j} = - \pi h_j.
\ee
For the interval $A$ in an infinite chain, one has
\be \label{hj1}
h_j^{\infty} = \f{j(\ell-j)}{\ell},
\ee
while in a periodic system of length $L$
\be \label{hj2}
h_j^{\rm PBC} = \f{\sin\f{\pi j}{L}\sin\f{\pi (\ell-j)}{L}}{\f{\pi}{L}\sin\f{\pi\ell}{L}}.
\ee
The corresponding matrix of the correlation functions of the BW modular Hamiltonian is
\be \label{C_HBW}
C^\BW = \f{1}{1+\ep^{H^\BW}}.
\ee


\subsection{One interval embedded in an infinite chain}

In this subsection we consider the ground state of an infinite XX chain.
By construction, the exact correlation matrix $C$ and the BW one $C^{\BW}$ commute \cite{eisler2017analytical}.
Hence the corresponding RDM also commute.
These commuting RDMs have the same eigenvectors, but may have different eigenvalues that we will use to
compute various distances and other related quantities.

\subsubsection{Trace distance and fidelity}

We calculate the trace distance, Schatten $n$-distance, and fidelity between the exact RDM and BW RDM.
We exploit the commutativity of the BW RDM with the exact one in our numerical calculations.
The $\ell\times\ell$ correlation matrix $C$ defined in (\ref{Cdef}) with (\ref{fjline}) has eigenvalues $\m_j$, $j=1,2,\cdots,\ell$, and in the diagonal basis the $2^\ell\times2^\ell$ RDM takes the form \cite{Vidal:2002rm,Latorre:2003kg}
\be \label{rhoAdiag}
\r_A = \bigotimes_{j=1}^\ell \bigg( \ba{cc}\m_j & \\ & 1-\m_j \ea \bigg).
\ee
As the correlation matrix $C$ commutes with the BW one $C^\BW$, which is defined in (\ref{C_HBW}) with (\ref{HBWjjp1}) and (\ref{hj1}),
they can be diagonalized simultaneously.
We denote the eigenvalues of the BW correlation matrix $C^\BW$ as $\n_j$, $j=1,2,\cdots,\ell$, and, in the same basis as (\ref{rhoAdiag}), the BW RDM is
\be
\r_A^\BW = \bigotimes_{j=1}^\ell \bigg( \ba{cc}\n_j & \\ & 1-\n_j \ea \bigg).
\ee
To calculate the trace distance $D(\r_A,\r_A^\BW)$ and general Schatten $n$-distance we need the explicit eigenvalues of the two RDMs.
Conversely, thanks to the commutativity of the RDMs $\r_A$, $\r_A^\BW$,
the fidelity is extracted by the simple formula
\be
F(\r_A,\r_A^\BW) = \tr_A \sr{\r_A\r_A^\BW} = \prod_{j=1}^\ell \Big[ \sqrt{\m_j\n_j} + \sqrt{(1-\m_j)(1-\n_j)} \Big],
\ee
that does not requires the reconstruction of the spectrum of the RDMs.
Similar equations can be found for all Schatten $n_e$-distance with $n_e$ being an even integer
\be
D_{n_e}( \r_A,\r_A^\BW ) = \f{1}{2^{1/n_e}} [ \tr_A( \r_A - \r_A^\BW )^{n_e} ]^{1/n_e}.
\ee
For example, the Schatten 2-distance is
\be
D_2(\r_A,\r_A^\BW) = \f{1}{2^{1/2}} \Big\{
    \prod_{j=1}^\ell \big[ \m_j^2 + (1-\m_j)^2 \big]
- 2 \prod_{j=1}^\ell \big[ \m_j \n_j + (1-\m_j) (1-\n_j) \big]
+   \prod_{j=1}^\ell \big[ \n_j^2 + (1-\n_j)^2 \big]
\Big\}^{1/2}.
\ee
As a consequence, the calculations of the fidelity and Schatten $n_e$-distances are much more effective because we only need to work
with the $\ell$ eigenvalues of the correlation matrix. Hence, we will access subsystem sizes up to $\ell\sim1000$ (but larger are also possible).
In contrast, for the trace distance and Schatten $n_o$-distances we have to reconstruct the $2^\ell$ eigenvalues of the RDMs;
for practical reasons we  cut off the eigenvalues smaller than $10^{-10}$, but also in this way we can go up at most to $\ell\sim100$.

\begin{figure}[t]
  \centering
  \includegraphics[height=0.3\textwidth]{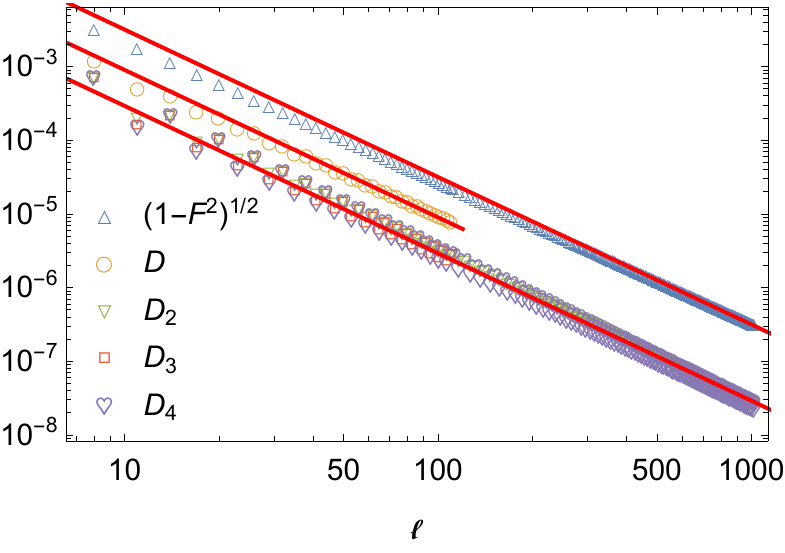}
  \caption{The trace distance $D \equiv D(\r_A,\r_A^\BW)$, Schatten $n$-distance $D_n \equiv D_n(\r_A,\r_A^\BW)$, and fidelity $F \equiv F(\r_A,\r_A^\BW)$ between the BW and the exact RDMs for one interval in an infinite XX chain.
  The fidelity provides an upper bound for the trace distance, i.e. $D\leq\sr{1-F^2}$.
  The symbols are the spin chain numerical results, and the solid lines are guides for the eyes going as $\ell^{-2}$.
 }
  \label{XXLineDis}
\end{figure}

The numerical results for the trace distance, Schatten $n$-distance, and fidelity are reported in Fig.~\ref{XXLineDis}.
From the figure it is evident that all of them decay algebraically as a function of  $\ell$, with an exponent that is compatible with $-2$ in all considered cases
(in spite of the relatively small $\ell$ that are accessible for the trace distance).
Hence, from now on we assume the result $D(\r_A,\r_A^\BW) \pp \ell^{-2}$, but we stress
that our conclusions do not change qualitatively as far as the decay is faster than $ \ell^{-1}$.


\subsubsection{Entanglement entropy}

By virtue of the Fannes-Audenaert inequality (\ref{FAI}), the behavior of the entanglement entropy is constrained by the trace distance.
Since we have found $D(\r_A,\r_A^\BW) \pp \ell^{-2}$, and since $d_A=2^\ell$, the Fannes-Audenaert inequality characterizes the difference of the entanglement entropies as
\be
| S_A - S_A^\BW | \lesssim \ell^{-1}.
\label{Sbou}
\ee
This is consistent with the results in \cite{Mendes-Santos:2019ine,Mendes-Santos:2019tmf}.
The inequalities for R\'enyi entropies (\ref{ieqSn1}) and (\ref{ieqSn2}) do not provide useful bounds;
anyhow their absence by no means implies that their differences are not vanishing as $\ell\to\infty$.

\begin{figure}[t]
  \centering
  \includegraphics[height=0.32\textwidth]{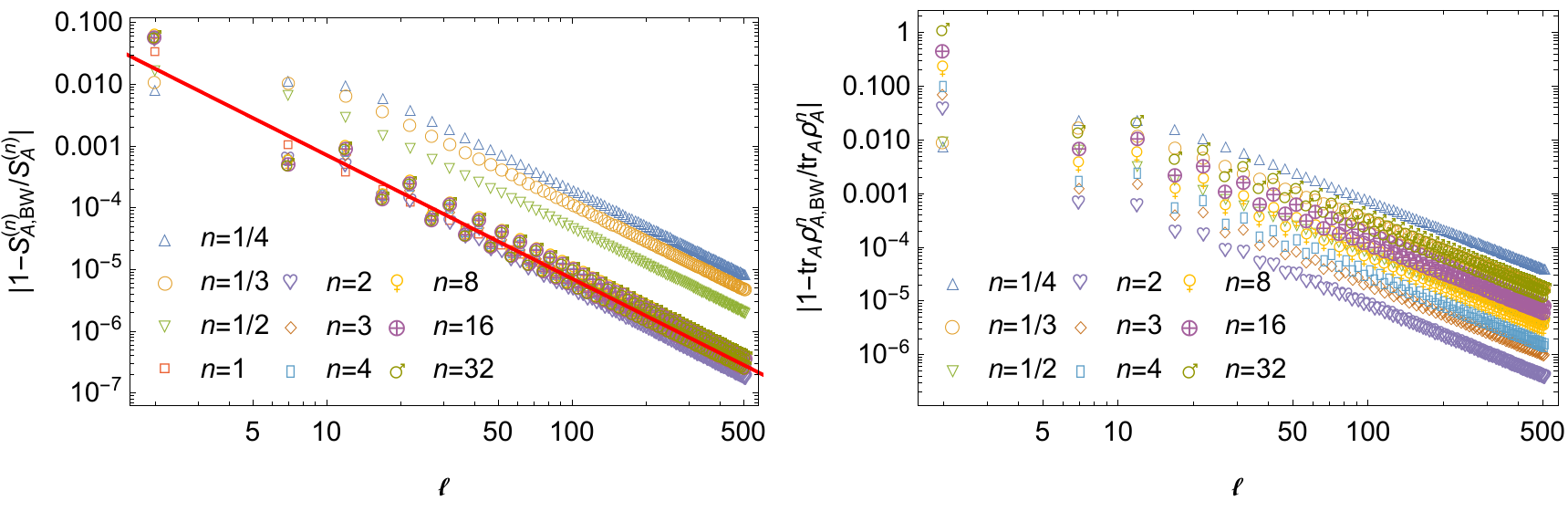}
\caption{Comparison of the R\'enyi entropies (left) and RDM moments (right) derived using the BW RDM
  with the exact ones for an interval of length $\ell$ in the ground state of an infinite XX chain.
  The symbols are the numerical data and the full line is a guide for the eyes, proportional to $\ell^{-2}$.
  Evidently, as $\ell\to\infty$, the BW RDM reproduces the exact results.}
  \label{XXLineEESnMn}
\end{figure}

We now present numerical computations for the differences of R\'enyi entropies  for several values of $n$, keeping
in mind the previous bound from Fannes-Audenaert inequality.
We calculate the entanglement entropy and R\'enyi entropy numerically from correlation matrix following \cite{Vidal:2002rm,Latorre:2003kg}.
The correlation matrix $C$ has eigenvalues $\m_j$ with $j=1,2,\cdots,\ell$, and the entanglement entropy, R\'enyi entropy, and RDM moment are
\bea
&& S_A = - \sum_{j=1}^\ell \big[ \m_j \log \m_j + (1-\m_j) \log (1-\m_j) \big], \nn\\
&& S_A^{(n)} = - \f{1}{n-1} \sum_{j=1}^\ell \log\big[ \m_j^n + (1-\m_j)^n \big], \nn\\
&& \tr_A\r_A^n = \prod_{j=1}^\ell \big[ \m_j^n + (1-\m_j)^n \big].
\eea
The same formulas are valid for the entanglement entropy, R\'enyi entropy, and RDM moment for the BW RDM just replacing
the eigenvalues $\m_j$ with those of the BW correlation matrix $C^\BW$, that we denoted by $\n_j$.
We emphasize that we need to set a high precision in numerical calculations of the R\'enyi entropy, especially when the index is in the range $0<n<1$.
The corresponding results are reported in the left panel of Fig.~\ref{XXLineEESnMn}.
The entropy difference decays with subsystem size, following approximately the law $| S_A - S_A^\BW | \pp \ell^{-2}$, showing that the
bound \eqref{Sbou} is not tight for the von Neumann entropy.
From the figure, it is also evident that also all R\'enyi entropies with arbitrary $n$ behave exactly in the same fashion, although there is no significant bound in this case.

Now we move to the moments of the RDM for which we have the bounds in Eqs.  (\ref{ieqTn1}) and (\ref{ieqTn2}).
Only the one for RDM moments with $n>1$ is potentially worth to consider:
\be
| \tr_A\r_A^n - \tr_A\r_{A,\BW}^n | \lesssim \ell^{-2}.
\ee
However as we mentioned in section \ref{secTraceDis} the moments themselves of the real RDM $\rho_A$ are decaying to zero as $\tr_A\r_A^n \propto \ell^{-\frac16 (n-\f1n)}$.
Hence it is more meaningful to look at the ratio for which the bound (\ref{ieqTn2}) leads to
\be
\Big| 1 - \f{\tr_A\r_{A,\BW}^n}{\tr_A\r_A^n} \Big| \lesssim \ell^{-2+\frac16 (n-\f1n)},
\label{Sbou3}
\ee
which decays to zero as long as $n< 12.08\dots$.
In Fig.~\ref{XXLineEESnMn} we see that all the BW moments converge to the exact ones for any $n$, also rather 
large. A fit of this decay including all available data points provides exponents that are between 1.8 and 1.9 for 
all values of $n$ considered. We note however that, since these fits do not consider possible subleading corrections or 
possible short-distance deviations due to strong-UV effects, these exponents shall be taken with a grain of salt.

\subsubsection{Two-point fermion correlation function}\label{subsubseccor}

In this subsection we consider the two-point correlation function of fermion operators, i.e. $\langle a_j^\dag a_k\rangle$ with $j,k=1,2,\cdots,\ell$.
These are the building blocks of the correlation matrix $C$ and, by means of Wick theorem, of all
local observables within $A$.
For these correlations, the inequality (\ref{inequality}) with $\cO=a_j^\dag a_k$ constrains their behavior (notice that
the maximum singular value  $s_{\max}(\cO)$  is of order one).
From the scaling of the trace distance $D(\r_A,\r_A^\BW) \pp \ell^{-2}$ and the inequality (\ref{inequality}), we get the bound for the difference of
all observables with $s_{\max}(\cO)$ of order one:
\be
|\lag \cO \rag - \lag \cO \rag_\BW| \lesssim \ell^{-2}.
\ee
When the two fermion operators are at a distance much smaller than $\ell$, $\lag a_j^\dag a_k \rag$ is a constant (in $\ell$), and so we conclude
\be \label{cf1}
| 1 - \lag a_j^\dag a_k \rag_\BW /  \lag a_j^\dag a_k \rag | \lesssim \ell^{-2}, \qquad {\rm if}\; |j-k|\ll \ell.
\ee
When the two fermion operators are instead at a distance proportional to $\ell$,
we have $\lag a_j^\dag a_{j+ \alpha \ell} \rag \sim \ell^{-1}$ (cf. Eq. \eqref{fjline}), and  so
\be \label{cf2}
| 1 - \lag a_j^\dag a_k  \rag_\BW /  \lag a_j^\dag a_k  \rag | \lesssim \ell^{-1} \qquad {\rm if}\; |j-k|\propto \ell.
\ee

\begin{figure}[thp]
  \centering
  \includegraphics[width=.98\textwidth]{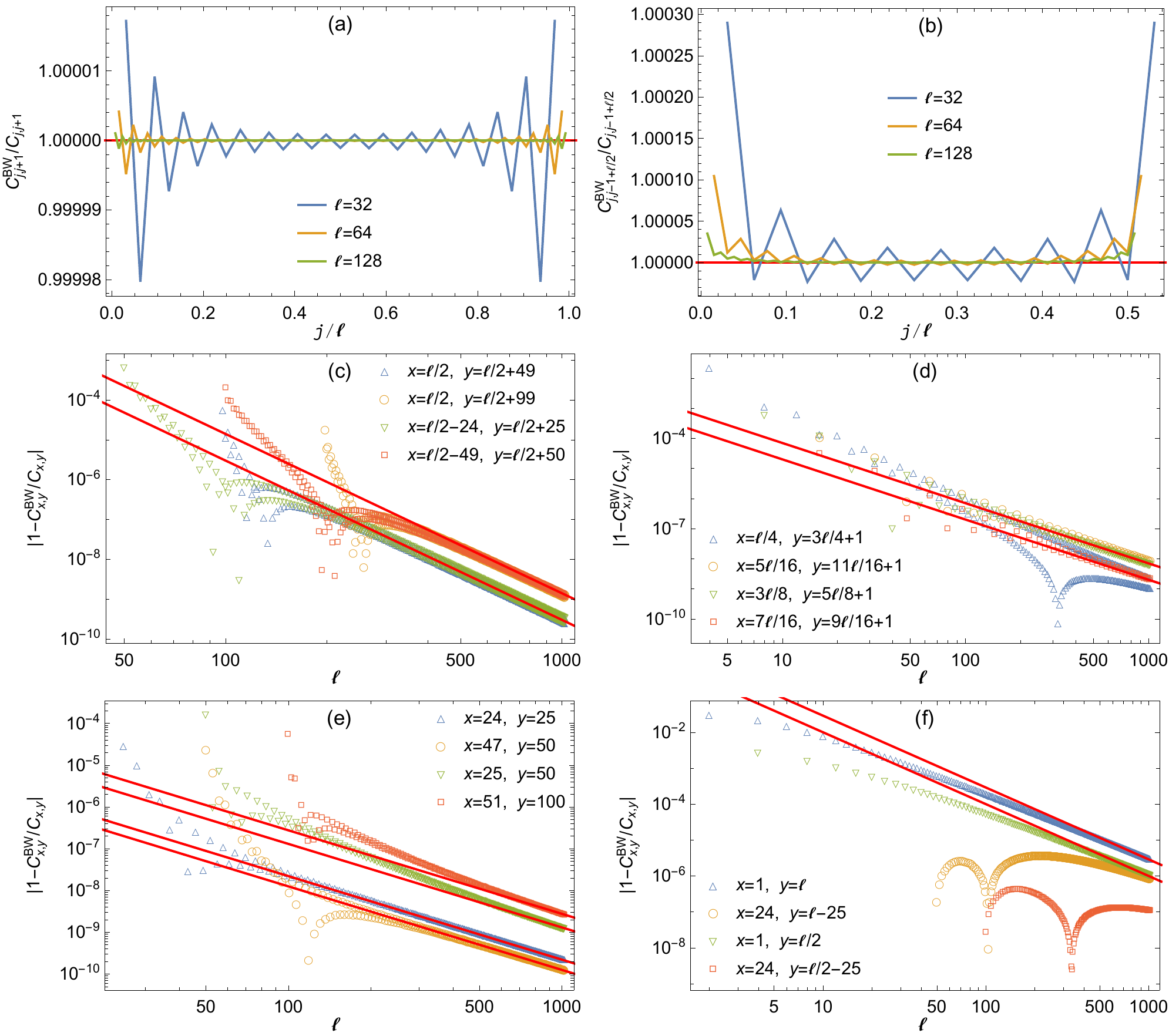}
  \caption{Comparison of exact correlation functions with those coming from the BW RDM for one interval of length $\ell$ in an infinite XX chain.
  Panels (a) and (b) show that  the BW  correlations  approach the exact ones restoring
  translational invariance as $\ell \to \inf$.
  In the panels (c)-to-(f) we report the relative difference of the correlations $|1- C^{\rm BW}_{x,y}/C_{x,y}|$.
  In (c), $x-y$ is finite and $x,y\propto \ell$.
  In (d) $x,y,x-y\propto \ell$.
  In (e) $x, y$ are both of order $1$.
  In (f) $x$ is $O(1)$, but $y\propto \ell$.
  In panel (c) the algebraic decay is proportional to $\ell^{-4}$ and in the other three cases to $\ell^{-2}$ (full lines in the various plots).
  }
  \label{XXLineCor}
\end{figure}

The comparison between the correlation functions from the BW RDM and the exact ones is shown in Fig.~\ref{XXLineCor}.
In the first row, we plot two examples at distance $1$ and $\ell/2-1$, as representative cases of small and large distances respectively.
As it is well known, we observe that the BW correlation function $C^{BW}$ breaks translational invariance.
However, such breaking is stronger at small $\ell$ and the symmetry is systematically restored as $\ell$ increases.
In the other four panels of Figure~\ref{XXLineCor} we compare the entries of the exact correlation matrix $C$ and the $C^\BW$, by
plotting their relative differences and studying the behavior  for large $\ell$.
There are two scales that matter: the distance from the boundary of the interval and the distances between the points.
Both of them can be either finite or growing with $\ell$ and so we have four possible cases.
In the figure we report four examples, one for each possible case.
It turns out that the differences for all cases decay like a power law, but the exponent depends on both the relevant scales mentioned above.
The numerical results are roughly compatible with the following picture:
In the large $\ell$ limit, the difference decays with an exponent in $\ell$ equal  (approximately) to  $-4$ or $-2$.
The decay power is about $-4$, only if both the following conditions are met:
\begin{enumerate}
  \item the distance of the two operators is finite, and
  \item the distances of the two operators from the boundaries of the interval are proportional to $\ell$.
\end{enumerate}
Otherwise, the exponent is approximately $-2$.
Let us see how these rules apply to the panel of Fig.~\ref{XXLineCor};
only in (c) both conditions are satisfied and the decay is indeed proportional to $\ell^{-4}$ (full line);
in all other panels, at most one condition is true (in (d) only (2), in (e) only (1), and in (f) none),
hence the decay is proportional to $\ell^{-2}$ (full lines in all panels).

Most of the cases that we considered follow clearly the above picture for the exponent.
However, the crossover between the two regimes strongly affects the data
for intermediate values of $\ell$, as it is evident from many curves in Fig.~\ref{XXLineCor}, but this is not surprising.
In Fig.~\ref{XXLineCor} there exist strong dips for some curves at intermediate values 
of $\ell$, which are due to changes of signs of the differences $1- C^{\rm BW}_{x,y}/C_{x,y}$.
Obviously all these observations about the numerical results are consistent with the bounds (\ref{cf1}) and (\ref{cf2}).

\subsubsection{Formation probabilities}
In this subsection we study non-local quantities called formation probabilities.
The most known of this kind of  observables is the emptiness formation probability (EFP) $P_\pm$ which has a long history, see \cite{Korepin:1994ui,Essler:1994se,Essler:1995vp,shiroishi2001emptiness,Kitanine2002a,Korepin2003,Franchini:2005uv,Stephan2013,najafi2016formation,Rajabpour2015,Viti2016,Rajabpour2016,Ares2020}.
$P_+$ ($P_-$) is the probability of having {\it all} the spins in a block of length $\ell$ pointing up (down) in the $\sigma^z_j$ basis.
More generally,  a  formation probability is defined as $P=\tr_A( \r_A |\psi\rag\lag\psi| )$, where $|\psi\rag$ is
a product state of the subsystem $A$ \cite{najafi2016formation,Najafi:2019ypm}.
Using $s_{\rm max}(|\psi\rag\lag\psi| )=1$ for product states and recalling that $D(\r_A,\r_A^\BW) \pp \ell^{-2}$, we get from (\ref{inequality}) that any formation probability should
satisfy
\be \label{ieqFP}
| P - P^\BW | \lesssim \ell^{-2},
\ee
but, as we shall see, this bound is not very meaningful because the FPs themselves are exponentially small.

\begin{figure}[t]
  \centering
  \includegraphics[height=0.32\textwidth]{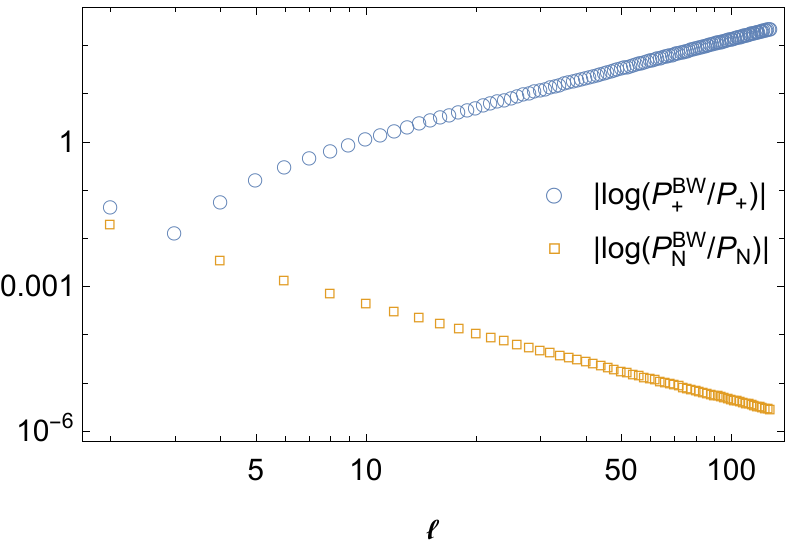}
  \caption{Comparison of the BW EFP and the BW NFP
  with the exact ones for one interval in the ground state of an infinite XX chain.
  While the NFP is correctly captured by the BW RDM, the same is not true for the EFP. }
\label{XXLineEFPNFP}
\end{figure}

Let us first analyse the emptiness formation probabilities; they are defined as
\be
P_\pm = \tr_A( \r_A \cO_\pm ), \qquad \cO_\pm = \prod_{j=1}^\ell \f{1\pm\s_j^z}{2}.
\ee
and can be rewritten as the determinant of an $\ell\times\ell$ matrix $S_\pm$ \cite{shiroishi2001emptiness,Franchini:2005uv}
\be \label{PpmSpm}
P_\pm = \det S_\pm, \qquad S^\pm_{j_1j_2} = \f12 ( \d_{j_1j_2} \pm \ii \lag d_{2j_1-1}d_{2j_2} \rag ).
\ee
The Majorana modes $d_m$ with $m=1,2,\cdots,2\ell$ are defined in (\ref{MajoranaModes}).
%
Inversion symmetry of the XX spin chain without transverse field 
guarantees $P_+=P_-$. 
In terms of the correlation matrix $C$, we have $S_+=1-C$ and $S_-=C$.
The known exact result for the large $\ell$ asymptotics of the EFP of the XX spin-chain in zero field is \cite{shiroishi2001emptiness}
\be
-\log P_+ = \f{\log2}{2} \ell^2 + \f14 \log\ell + \cdots,
\label{P+}
\ee
where the dots are subleading terms in the large $\ell$ limit.
Conversely, analyzing the numerical data for the EFP from the BW modular Hamiltonian, we get a very accurate fit with the form%
\footnote{This leading order result $\f{1}{3} \ell^2$ can be obtained analytically by integrating the spectrum of the BW RDM obtained in Ref. \cite{Slepian1978Prolate,eisler2017analytical}, as suggested to us by Viktor Eisler.}
\be
-\log P_+^\BW = \f{1}{3} \ell^2 + \f13 \log\ell + \cdots.
\label{P+bw}
\ee
The difference $|\log P_+-\log P_+^\BW|$ is shown in figure~\ref{XXLineEFPNFP} and grows for large $\ell$ as $\ell^2$, compatible with
the forms \eqref{P+} and \eqref{P+bw}. This behavior will be discussed at the end of the subsection.

We now consider the N\'eel formation probability (NFP) $P_\rN$, i.e. the formation probability of the N\'eel state.
This can be written in terms of the matrix $S_+$ in (\ref{PpmSpm}) as \cite{najafi2016formation,Najafi:2019ypm}
\be
P_\rN = ( \det S_+ ) \Big(  {\rm sdet} \f{1-S_+}{S_+} \Big),
\ee
where ${\rm sdet} (M)$ stands for the determinant of the submatrix of $M$ constructed in such a way that all rows and columns with even (or odd) index are removed.
The numerical calculations
presented in Ref. \cite{najafi2016formation} suggest the scaling
\be
-\log P_\rN = \f{\log2}{2} \ell + \f18 \log\ell + \cdots,
\ee
where the dots stand for subleading terms.
We find that the logarithmic NFP derived by using the BW modular Hamiltonian is compatible with this ansatz:
in figure~\ref{XXLineEFPNFP} we report the difference between the logarithms of the exact and BW NFP, showing that indeed it goes to zero
for large $\ell$.

The EFP and NFP themselves are exponentially small as $\ell \to \inf$, and so the inequality (\ref{ieqFP}) are satisfied trivially.
The decaying trace distance for large subsystems guarantees the decaying differences of formation probabilities, but the difference of a particular formation probability may not approach each other in $\ell \to \inf$ limit.
This is similar to our conclusion in section~\ref{sectoy}, i.e.\ that two closed RDMs may have very distinct modular Hamiltonians.
For this reason, it is not at all surprising that the EFP is not captured by the BW RDM; quite oppositely, we find  remarkable and unexpected that the NFP is well
described by BW.

\subsection{One interval in a finite periodic chain}

\begin{figure}[t]
  \centering
  \includegraphics[height=0.32\textwidth]{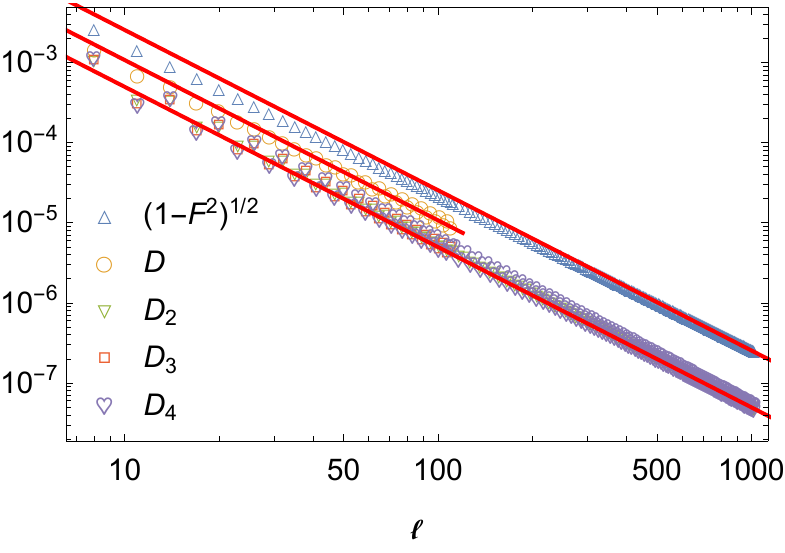}
  \caption{The trace distance, Schatten $n$-distance, and fidelity between the BW and the exact RDMs for one interval in a periodic XX chain.
  The empty plotmakers are spin chain numerical results, and the solid lines are guide for eyes decaying as $\ell^{-2}$.
  We fixed the ratio of the length of the interval $\ell$ and the length of the circle $L$ as $\ell={L}/{4}$.}\label{XXCircleDis}
\end{figure}

In this subsection, we move our attention to a finite XX chain of length $L$ with periodic boundary conditions.
For conciseness we focus on a subsystem of length $\ell=L/4$, but any other ratio of $\ell/L$ would work the same.
In analogy to the study of the infinite system, we first consider the behavior of trace distance, Schatten $n$-distance, and fidelity between the exact RDM $\rho_A$ and the BW RDM, i.e. $\rho_A^{\BW}$, with the modular Hamiltonian defined in (\ref{HABWdef}) with (\ref{HBWjjp1}) and (\ref{hj2}).
In our calculations, we rely on the fact that the exact correlation matrix $C$ (\ref{Cdef}) calculated using the equation  (\ref{fjcircle}) commutes with the matrix $H^\BW$ (\ref{HBWjjp1}) calculated by (\ref{hj2}), see Ref. \cite{eisler2018properties}.
Our results are shown in Fig.~\ref{XXCircleDis}; the trace distance, the Schatten $n$-distance, and $\sr{1-F^2}$ with the fidelity $F$ decay all  as approximately $\ell^{-2}$.
It is worth mentioning that we find similar results also for open boundary conditions.

\begin{figure}[t]
  \centering
  \includegraphics[height=0.32\textwidth]{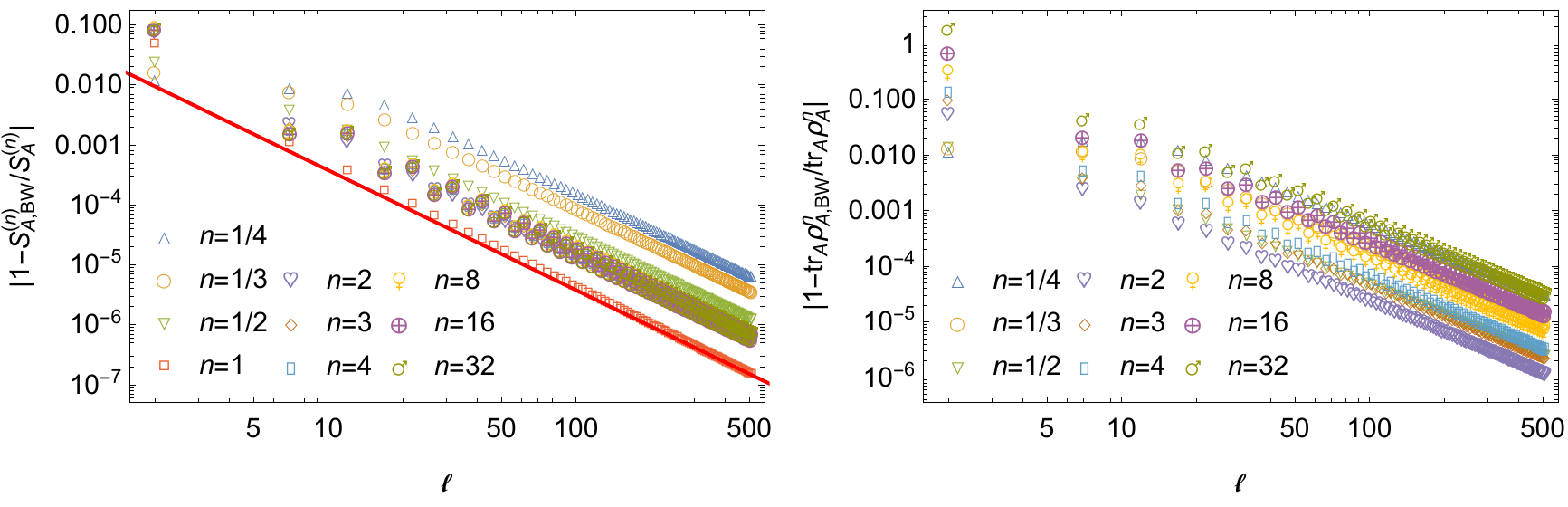}
  \caption{Comparison of the entanglement entropy, R\'enyi entropies, and RDM moments derived using
  the BW RDM with the exact ones for one interval in a periodic XX chain with zero magnetic field and for $L=4\ell$.}\label{XXCircleEESnMn}
\end{figure}

We now consider the behavior of the differences of the entanglement entropy, R\'enyi entropies, and RDM moments in figure~\ref{XXCircleEESnMn}.
We see that approximately $| S_A - S_A^\BW | \sim \ell^{-2}$. As in the previous subsection, the BW RDM can reproduce the correct R\'enyi entropy and RDM moments.
We also compare the correlation functions of fermion operators calculated using the BW RDM with the exact ones for one interval on a circle in the ground state of the XX chain. The correlation functions from the BW RDM approach the exact ones for large intervals. The results are very similar to Fig.~\ref{XXLineCor} and we will not show them here.

\section{The critical Ising chain}\label{secIsing}

The transverse field Ising  chain is defined by the  Hamiltonian
\be
H_{\rm Ising} = - \f12 \sum_{j=1}^L \big( \s_j^x\s_{j+1}^x + \l \s_j^z \big).
\ee
Its phase diagram includes a critical point at $\l=1$, on which we focus here.
The continuum limit of the critical Ising spin chain is a free massless Majorana fermion, which is a 2D CFT with the central charge $c=\f12$.
We recall that, in the massive regime, the entanglement Hamiltonian of the half-chain in the thermodynamic limit is known to correspond exactly
to the lattice version of the BW modular Hamiltonian~\cite{itoyama1987lattice}.
The Hamiltonian of the Ising spin chain can be diagonalized exactly by a Jordan-Wigner transformation (\ref{JWtransf}) and
moving to Majorana modes (\ref{MajoranaModes}) as for the XX spin chain.
We anticipate that most of the results we obtained for the XX chain will also apply here at the qualitative level;
as such, we will keep the discussion of in the form of a short summary, eventually emphasizing differences with respect to the aforementioned case.

Due to the fact that $U(1)$ magnetization conservation is replaced here by a global $\mathbb{Z}_2$ symmetry, we work directly in the Majorana basis $d_m$ with $m=1,2,\cdots,2\ell$ defined by Eq.~(\ref{MajoranaModes}). The exact modular Hamiltonian of a length $\ell$ interval takes the form
\be
H_A = \f12 \sum_{m_1,m_2=1}^{2\ell} W_{m_1m_2}d_{m_1} d_{m_2}.
\ee
The correlation matrix $\Gamma$ of Majorana operators is defined as
\be \label{Gdef1}
\G_{m_1m_2} = \lag d_{m_1} d_{m_2} \rag - \d_{m_1m_2}.
\ee
In the ground state of the critical Ising spin chain, the non-vanishing entries of the correlation matrix satisfy
\be \label{Gdef2}
\G_{2j_1-1,2j_2} = - \G_{2j_2,2j_1-1} = g_{j_2-j_1}.
\ee
For the interval embedded in the ground state of an infinite chain we have 
\be \label{gjline}
g_j^{\infty} = - \f{\ii}{\pi} \f{1}{j + \f12},
\ee
while on a periodic system of length $L$
\be \label{gjcircle}
g_j^{\rm PBC} = - \f{\ii}{L\sin[(j + \f12)\f{\pi}{L}]}.
\ee
The matrices $W$ and $\G$ are related as\cite{Fagotti:2010yr}
\be
\G = \tanh W.
\ee
The BW modular Hamiltonian of the critical Ising spin chain is
\be \label{HABWIsingDef}
H_A^\BW = \f12 \sum_{m_1,m_2=1}^{2\ell} W_{m_1m_2}^\BW d_{m_1} d_{m_2},
\ee
with the nonvanishing entries of the antisymmetric matrix
\bea \label{W2jm12jW2j2jp1}
&& W_{2j-1,2j}^\BW = - W_{2j,2j-1}^\BW = - \pi \ii h_{j-\f12}, ~~ j = 1,2,\cdots,\ell\nn\\
&& W_{2j,2j+1}^\BW = - W_{2j+1,2j}^\BW = - \pi \ii h_{j}, ~~ j = 1,2,\cdots,\ell-1.
\eea
The function $h_j$ is defined as (\ref{hj1}) or (\ref{hj2}), depending on the interval of interest.
The corresponding correlation matrix of the BW modular Hamiltonian is
\be \label{GBWdef}
\G^\BW = \tanh W^\BW.
\ee

\subsection{One interval in an infinite chain}

In this subsection we consider the ground state of an infinite chain.
The BW modular Hamiltonian is defined as (\ref{HABWIsingDef}) with (\ref{W2jm12jW2j2jp1}) and (\ref{hj1}).

\subsubsection{Trace distance and fidelity}

\begin{figure}[t]
  \centering
  \includegraphics[height=0.32\textwidth]{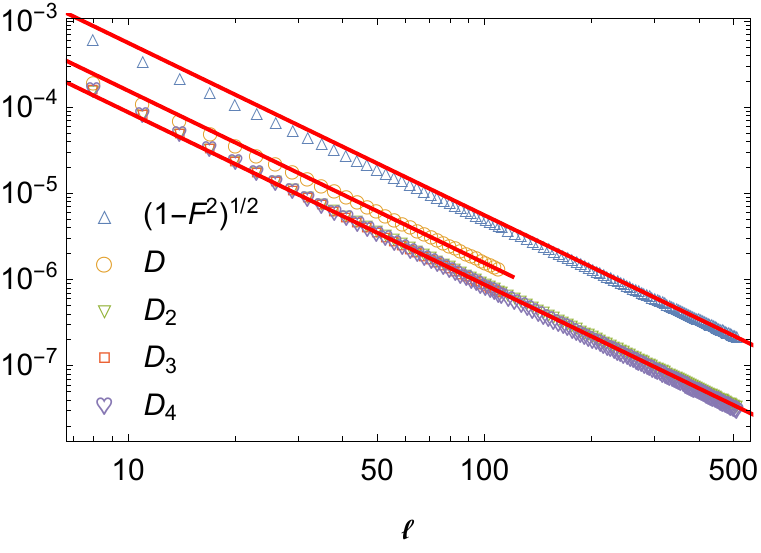}
  \caption{The trace distance, Schatten $n$-distance, and fidelity of the BW and the exact RDMs for one interval
  on an infinite chain in the ground state of the critical Ising spin chain.
  The empty plotmakers are spin chain numerical results. The solid lines are guide to the eyes going like $\ell^{-2}$.}\label{IsingLineDis}
\end{figure}

The exact correlation matrix $\G$ (\ref{Gdef1}) with (\ref{Gdef2}) and (\ref{gjline}) and the matrix $W^\BW$ (\ref{W2jm12jW2j2jp1}) with (\ref{hj1}) commute.
The exact correlation matrix $\G$ has eigenvalues $\pm\g_j$ with $j=1,2,\cdots,\ell$, and the BW correlation matrix $\G^\BW$ (\ref{GBWdef}) can be diagonalized under the same basis with the eigenvalues $\pm\d_j$ with $j=1,2,\cdots,\ell$.
In the same  basis, the exact and the BW RDM can be written respectively in the diagonal forms
\be
\r_A = \bigotimes_{j=1}^\ell \bigg( \ba{cc} \f{1+\g_j}{2} & \\ & \f{1-\g_j}{2} \ea \bigg), \qquad
\r_A^\BW = \bigotimes_{j=1}^\ell \bigg( \ba{cc} \f{1+\d_j}{2} & \\ & \f{1-\d_j}{2} \ea \bigg).
\ee
To calculate the trace distance and Schatten $n_o$-distance with $n_o$ being an odd integer we need the explicit eigenvalues of the RDMs, while we have simpler formulas to calculate the Schatten $n_e$-distance with $n_e$ being an even integer and fidelity.
As for the XX chain, we discard the eigenvalues of the RDMs smaller than $10^{-10}$, and this allow us to calculate the trace distance and Schatten $n_o$-distance up to $\ell\sim100$. However, this limitation does not apply to  fidelity and Schatten $n_e$-distance for which we go up to $\ell\sim500$.
The trace distance, Schatten distance, and the fidelity of the exact RDM and the BW RDM are shown in the figure~\ref{IsingLineDis}.
Similarly to the XX chain, the BW RDM becomes an increasingly better approximation of the exact RDM as the subsystem size $\ell \to \inf$.
By numerical fitting we get that the trace distance, Schatten distance and $\sr{1-F^2}$ with $F$ being the fidelity all decay approximately as $\ell^{-2}$.

\subsubsection{Entanglement entropy}

\begin{figure}[tbp]
  \centering
  \includegraphics[height=0.32\textwidth]{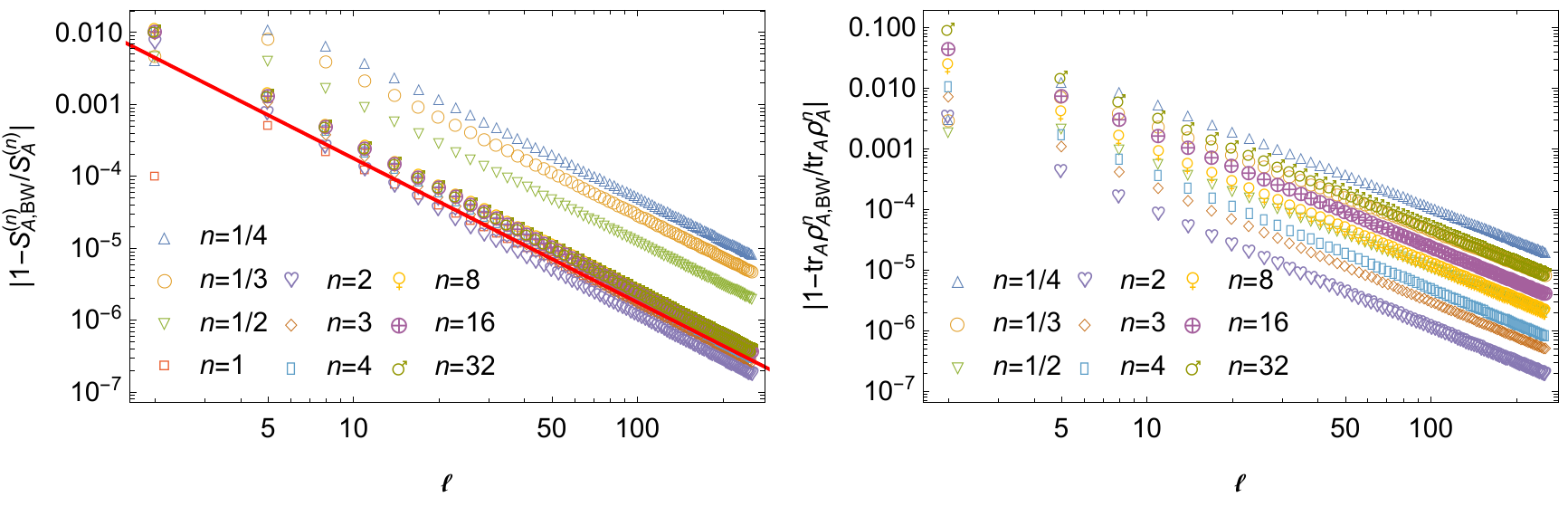}
  \caption{Comparison of the entanglement entropy, R\'enyi entropies,
  and RDM moments derived using the BW RDM with the exact ones
  for one interval on an infinite chain in the ground state of the critical Ising chain.
  The full line on the left is proportional to $\ell^{-2}$.}
  \label{IsingLineEESnMn}
\end{figure}

In this subsection we study the behavior of the differences of the entanglement entropy, R\'enyi entropies, and the
RDM moments in figure~\ref{IsingLineEESnMn}.
We see that approximately $|S_A - S_A^\BW| \pp \ell^{-2}$.
It is consistent with the results in \cite{Mendes-Santos:2019tmf}.
Again, similarly to the XX case, the BW RDM reproduces all the exact R\'enyi entropies and RDM moments.
The difference of R\'enyi entropies decay as $\ell^{-2}$, while the difference of the moments decay with an exponent
slightly smaller than $2$, in full analogy with the XX chain.

We mention that we tested also the scaling of the correlation functions.
All results are identical to the ones of the XX chain in Figure \ref{XXLineCor} and so we do not discuss them here.

\subsubsection{Formation probabilities}

\begin{figure}[t]
  \centering
    \includegraphics[height=0.32\textwidth]{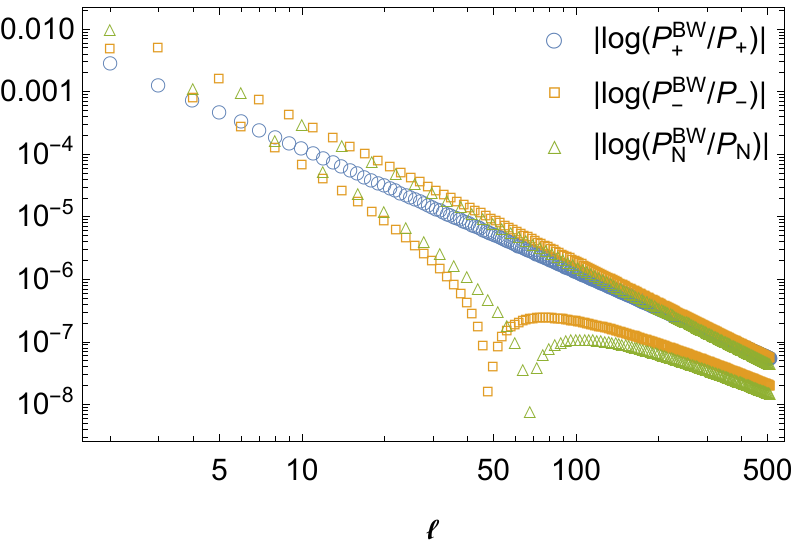}
  \caption{Comparison of the logarithmic EFPs and logarithmic NFP coming
  from the BW RDM with the exact ones for one interval on an infinite chain in the ground state of the critical Ising chain.
  Notice that $P_-$ and $P_N$ displays even-odd effects.}
  \label{IsingLineEFPNFP}
\end{figure}

In the critical Ising chain, the EFPs $P_\pm$ for all the spins up and all the spins down in the basis of $\s_j^z$ are different.
For the critical Ising spin chain, there are the exact results \cite{Franchini:2005uv}
\be \label{mlogPpmIsing}
-\log P_\pm = \Big( \log2 \mp \f{2G}{\pi} \Big) \ell + \f1{16} \log\ell + \cdots,
\ee
where $G$ is the Catalan's constant.
Note that $P_+>P_-$, i.e. the configuration with all the spins up (in the transverse direction) is preferred to the one with spin down,
as obvious from energetic arguments.
In terms of the correlation matrix $\G$ (\ref{Gdef1}) with (\ref{Gdef2}) and (\ref{gjline}), we have \cite{shiroishi2001emptiness,Franchini:2005uv}
\be
P_\pm = \det S_\pm, \qquad S^\pm_{j_1j_2} = \f12 ( \d_{j_1j_2} \pm \ii \G_{2j_1-1,2j_2} ).
\ee
The same formula works also for the BW correlation matrix $\G^\BW$ (\ref{GBWdef}) with (\ref{W2jm12jW2j2jp1}) and (\ref{hj1}) to get the
EFP $P^{BW}_{\pm}$. The numerical results are shown in figure~\ref{IsingLineEFPNFP}.
Unlike the case in the XX spin chain, the BW modular Hamiltonian in the Ising spin chain perfectly reproduces the EFPs.

In the critical Ising spin chain numerical calculations suggest that for the NFP we have \cite{najafi2016formation}:
\be
-\log P_\rN \approx 0.985 \ell +\f{1}{16} \log\ell + \cdots.
\ee
We find NFP coming from the BW modular Hamiltonian also matches this result when the subsystem size is large, as shown in figure~\ref{IsingLineEFPNFP}.

\subsection{One interval in a finite periodic chain}

\begin{figure}[tbp]
  \centering
  \includegraphics[height=0.32\textwidth]{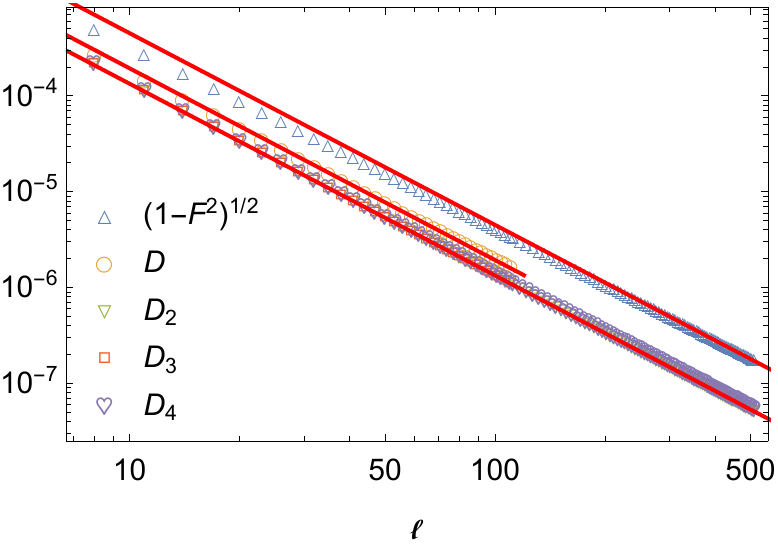}
  \caption{The trace distance, Schatten $n$-distance, and fidelity of the BW and exact RDMs for
  one interval  in the ground state of  a periodic critical Ising spin chain.
  The empty plotmakers are spin chain numerical results, and the solid lines are decaying as $\ell^{-2}$.
  We have fixed the ratio of the length of the interval $\ell$ and the length of the circle $L$ as $\ell={L}/{4}$.}\label{IsingCircleDis}
\end{figure}

In this subsection we consider the ground state of a periodic system of length $L=4\ell$.
The exact correlation matrix $\G$ (\ref{Gdef1}) with (\ref{Gdef2}) and (\ref{gjcircle}) commute with $W^\BW$ (\ref{W2jm12jW2j2jp1}) with (\ref{hj2}).
The trace distance, Schatten distance and fidelity of the exact RDM and BW RDM are shown in figure~\ref{IsingCircleDis}.
The trace distance, Schatten distance and $\sr{1-F^2}$ with $F$ being the fidelity all decay approximately as $\ell^{-2}$.

\begin{figure}[tbp]
  \centering
  \includegraphics[height=0.32\textwidth]{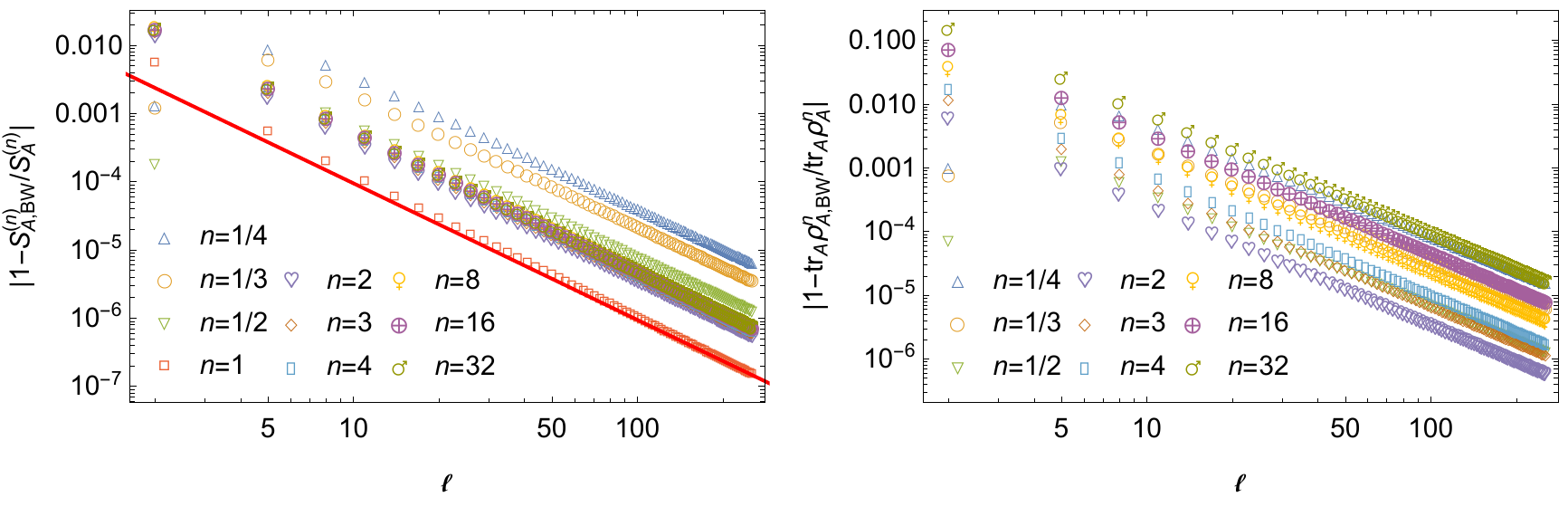}
  \caption{Comparison of the entanglement entropy, R\'enyi entropies, and RDM moments coming
  from the BW RDM with the exact ones for one interval on a circle in the ground state of the critical Ising chain.}\label{IsingCircleEESnMn}
\end{figure}

We show the behavior of the differences of the entanglement entropy, R\'enyi entropies, and RDM moments in Figure~\ref{IsingCircleEESnMn}.
We see that approximately $| S_A - S_A^\BW | \pp \ell^{-2}$.
The BW RDM reproduces the exact entanglement entropy, R\'enyi entropies and RDM moments for large subsystem sizes.

\section{Conclusion}\label{seccon}

We compared the reduced density matrices coming from the discretization of the BW
modular Hamiltonian and the exact ones for the ground state of the
XX chain with a zero magnetic field and the critical transverse field Ising chain.
In order to provide a meaningful metric comparison between density matrices, we have considered the trace distance between
the exact RDM and the BW RDM as our main figure of merit.
We found that the trace distance between BW RDMs and the exact ones goes to zero approximately as $\ell^{-2}$ when the length of
the interval $\ell$ goes to infinity.
The behavior of the trace distance gives strong constraints on the entanglement entropy and correlation functions.
The differences of the entanglement entropy and correlation functions also go to zero as $\ell \to \inf$.

We have found that the R\'enyi entropies and moments of any order are all well reproduced by the BW RDM.
Their differences decay algebraically for large $\ell$, although in most cases the known sharp bounds provides
no constraint at all. It is not known yet what special properties of the ground state of critical system enforce such
peculiar scaling, but it is clearly worth investigating because it is related to the working accuracy of the BW RDM.
However, we found that the most peculiar behavior is displayed by the formation probabilities.
Indeed, in the ground state they decay exponentially fast with the length of the interval and so, a priori, we would not
have expected to be well reproduced by the BW approximation.
Instead, only the emptiness formation probability of the XX chain is not captured by the BW RDM, while the
N\'eel formation probability in XX and all the studied ones in Ising are, surprisingly, captured by this approximation.
We do not know whether and how this could be connected with the Gaussian vanishing for large $\ell$ of the EFP
in the XX chain, while all the others are simple exponentials.

\section*{Acknowledgements}

We thank V.~Eisler for helpful discussions and reading the manuscript.
MD and MAR thank G. Giudici and T. Mendes-Santos for discussions and collaborations on related topics.
MAR also thanks ICTP and SISSA for hospitality during the initial and final stages of this work.


\paragraph{Funding information}
PC and JZ acknowledge support from ERC under Consolidator grant number 771536 (NEMO).
MD acknowledges support ERC under grant number 758329 (AGEnTh), and the European Union's Horizon 2020 research and innovation programme under grant agreement No 817482 (Pasquans).
MAR thanks CNPq and FAPERJ (grant number 210.354/2018) for partial support.

\providecommand{\href}[2]{#2}\begingroup\raggedright\endgroup

\nolinenumbers


\begin{thebibliography}{10}

\bibitem{Amico:2007ag}
L.~Amico, R.~Fazio, A.~Osterloh and V.~Vedral, \textit{{Entanglement in
  many-body systems}}, \href{http://dx.doi.org/10.1103/RevModPhys.80.517}{Rev.
  Mod. Phys. {\bfseries 80}, 517 (2008)},
  [\href{https://arxiv.org/abs/quant-ph/0703044}{{\ttfamily
  arXiv:quant-ph/0703044}}].

\bibitem{Eisert:2008ur}
J.~Eisert, M.~Cramer and M.~B. Plenio, \textit{{Area laws for the entanglement
  entropy - a review}}, \href{http://dx.doi.org/10.1103/RevModPhys.82.277}{Rev.
  Mod. Phys. {\bfseries 82}, 277 (2010)},
  [\href{https://arxiv.org/abs/0808.3773}{{\ttfamily arXiv:0808.3773}}].

\bibitem{calabrese2009entanglement}
P.~Calabrese, J.~Cardy and B.~Doyon, \textit{{Entanglement entropy in extended
  quantum systems}}, \href{http://dx.doi.org/10.1088/1751-8121/42/50/500301}{J.
  Phys. A {\bfseries 42}, 500301 (2009)}.

\bibitem{Laflorencie:2015eck}
N.~Laflorencie, \textit{{Quantum entanglement in condensed matter systems}},
  \href{http://dx.doi.org/10.1016/j.physrep.2016.06.008}{Phys. Rept. {\bfseries
  646}, 1 (2016)}, [\href{https://arxiv.org/abs/1512.03388}{{\ttfamily
  arXiv:1512.03388}}].

\bibitem{Witten:2018lha}
E.~Witten, \textit{{APS Medal for Exceptional Achievement in Research: Invited
  article on entanglement properties of quantum field theory}},
  \href{http://dx.doi.org/10.1103/RevModPhys.90.045003}{Rev. Mod. Phys.
  {\bfseries 90}, 045003 (2018)},
  [\href{https://arxiv.org/abs/1803.04993}{{\ttfamily arXiv:1803.04993}}].




\bibitem{daley2012measuring}
A.~Daley, H.~Pichler, J.~Schachenmayer and P.~Zoller, \textit{{Measuring
  entanglement growth in quench dynamics of bosons in an optical lattice}},
  \href{http://dx.doi.org/10.1103/PhysRevLett.109.020505}{Phys. Rev. Lett.
  {\bfseries 109}, 020505 (2012)},
  [\href{https://arxiv.org/abs/1205.1521}{{\ttfamily arXiv:1205.1521}}].

\bibitem{islam2015measuring}
R.~Islam, R.~Ma, P.~M. Preiss, M.~E. Tai, A.~Lukin, M.~Rispoli, and  M. Greiner,
  \textit{Measuring entanglement entropy in a quantum many-body system},
  \href{http://dx.doi.org/10.1038/nature15750}{Nature {\bfseries 528}, 77
  (2015)}, [\href{https://arxiv.org/abs/1509.01160}{{\ttfamily
  arXiv:1509.01160}}].

\bibitem{tta}
A. M. Kaufman, M. E. Tai, A. Lukin, M. Rispoli, R. Schittko, P. M. Preiss, and M. Greiner,
{\it Quantum thermalization through entanglement in an isolated many-body system}, \href{http://dx.doi.org/10.1126/science.aaf6725}{Science {\bf 353},  794 (2016)}, [\href{https://arxiv.org/abs/1603.04409}{{\ttfamily arXiv:1603.04409}}].

\bibitem{elben2018renyi}
A.~Elben, B.~Vermersch, M.~Dalmonte, J.~I. Cirac and P.~Zoller,
  \textit{{R{\'e}nyi Entropies from Random Quenches in Atomic Hubbard and Spin
  Models}}, \href{http://dx.doi.org/10.1103/PhysRevLett.120.050406}{Phys. Rev.
  Lett. {\bfseries 120}, 050406 (2018)},
  [\href{https://arxiv.org/abs/1709.05060}{{\ttfamily arXiv:1709.05060}}].

\bibitem{vermersch2018unitary}
B.~Vermersch, A.~Elben, M.~Dalmonte, J.~I. Cirac and P.~Zoller,
  \textit{{Unitary n-designs via random quenches in atomic Hubbard and spin
  models: Application to the measurement of R{\'e}nyi entropies}},
  \href{http://dx.doi.org/10.1103/PhysRevA.97.023604}{Phys. Rev. A {\bfseries
  97}, 023604 (2018)}, [\href{https://arxiv.org/abs/1801.00999}{{\ttfamily
  arXiv:1801.00999}}].

\bibitem{brydges2019probing}
T.~Brydges, A.~Elben, P.~Jurcevic, B.~Vermersch, C.~Maier, B.~P. Lanyon, P. Zoller, R. Blatt, and C. F. Roos,
  \textit{{Probing R{\'e}nyi entanglement entropy via randomized
  measurements}}, \href{http://dx.doi.org/10.1126/science.aau4963}{Science
  {\bfseries 364}, 260 (2019)},
  [\href{https://arxiv.org/abs/1806.05747}{{\ttfamily arXiv:1806.05747}}].




\bibitem{li2008entanglement}
H.~Li and F.~D.~M. Haldane, \textit{{Entanglement spectrum as a generalization
  of entanglement entropy: Identification of topological order in non-abelian
  fractional quantum Hall effect states}},
  \href{http://dx.doi.org/10.1103/PhysRevLett.101.010504}{Phys. Rev. Lett.
  {\bfseries 101}, 010504 (2008)},
  [\href{https://arxiv.org/abs/0805.0332}{{\ttfamily arXiv:0805.0332}}].

\bibitem{Bisognano:1975ih}
J.~J. Bisognano and E.~H. Wichmann, \textit{{On the Duality Condition for a
  Hermitian Scalar Field}}, \href{http://dx.doi.org/10.1063/1.522605}{J. Math.
  Phys. {\bfseries 16}, 985--1007 (1975)}.

\bibitem{Bisognano:1976za}
J.~J. Bisognano and E.~H. Wichmann, \textit{{On the Duality Condition for
  Quantum Fields}}, \href{http://dx.doi.org/10.1063/1.522898}{J. Math. Phys.
  {\bfseries 17}, 303 (1976)}.

\bibitem{itoyama1987lattice}
H.~Itoyama and H.~B. Thacker, \textit{{Lattice Virasoro algebra and corner
  transfer matrices in the Baxter eight-vertex model}},
  \href{http://dx.doi.org/10.1103/PhysRevLett.58.1395}{Phys. Rev. Lett.
  {\bfseries 58}, 1395 (1987)}.

\bibitem{peschel1999density}
I.~Peschel, M.~Kaulke and {\"O}.~Legeza, \textit{Density-matrix spectra for
  integrable models},
  \href{http://dx.doi.org/10.1002/(SICI)1521-3889(199902)8:2<153::AID-ANDP153>3.0.CO;2-N}{Ann.
  der Phys. {\bfseries 8}, 153--164 (1999)},
  [\href{https://arxiv.org/abs/cond-mat/9810174}{{\ttfamily
  arXiv:cond-mat/9810174}}].

\bibitem{nienhuis2009entanglement}
B.~Nienhuis, M.~Campostrini and P.~Calabrese, \textit{Entanglement,
  combinatorics and finite-size effects in spin chains},
  \href{http://dx.doi.org/10.1088/1742-5468/2009/02/P02063}{J. Stat. Mech.
  (2009) P02063}, [\href{https://arxiv.org/abs/0808.2741}{{\ttfamily
  arXiv:0808.2741}}].

\bibitem{peschel2009reduced}
I.~Peschel and V.~Eisler, \textit{Reduced density matrices and entanglement
  entropy in free lattice models},
  \href{http://dx.doi.org/10.1088/1751-8113/42/50/504003}{J. Phys. A {\bfseries 42}, 504003 (2009)},
  [\href{https://arxiv.org/abs/0906.1663}{{\ttfamily arXiv:0906.1663}}].

\bibitem{eisler2017analytical}
V.~Eisler and I.~Peschel, \textit{{Analytical results for the entanglement
  Hamiltonian of a free-fermion chain}},
  \href{http://dx.doi.org/10.1088/1751-8121/aa76b5}{J. Phys. A  {\bf 50}, 284003 (2017)},
  [\href{https://arxiv.org/abs/1703.08126}{{\ttfamily arXiv:1703.08126}}].

\bibitem{eisler2018properties}
V.~Eisler and I.~Peschel, \textit{{Properties of the entanglement Hamiltonian
  for finite free-fermion chains}},
  \href{http://dx.doi.org/10.1088/1742-5468/aace2b}{J. Stat. Mech. (2018)
  104001}, [\href{https://arxiv.org/abs/1805.00078}{{\ttfamily
  arXiv:1805.00078}}].

\bibitem{Eisler:2019rnr}
V.~Eisler, E.~Tonni and I.~Peschel, \textit{{On the continuum limit of the
  entanglement Hamiltonian}},
  \href{http://dx.doi.org/10.1088/1742-5468/ab1f0e}{J. Stat. Mech. (2019)
  073101}, [\href{https://arxiv.org/abs/1902.04474}{{\ttfamily
  arXiv:1902.04474}}].

\bibitem{DiGiulio:2019cxv}
G.~Di~Giulio and E.~Tonni, \textit{{On entanglement hamiltonians of an interval
  in massless harmonic chains}},
  \href{https://arxiv.org/abs/1911.07188}{{\ttfamily arXiv:1911.07188}}.

\bibitem{Dalmonte:2017bzm}
M.~Dalmonte, B.~Vermersch and P.~Zoller, \textit{{Quantum Simulation and
  Spectroscopy of Entanglement Hamiltonians}},
  \href{http://dx.doi.org/10.1038/s41567-018-0151-7}{Nat. Phys. {\bfseries 14},
  827 (2018)}, [\href{https://arxiv.org/abs/1707.04455}{{\ttfamily
  arXiv:1707.04455}}].

\bibitem{Giudici:2018izb}
G.~Giudici, T.~Mendes-Santos, P.~Calabrese and M.~Dalmonte,
  \textit{{Entanglement Hamiltonians of lattice models via the
  Bisognano-Wichmann theorem}},
  \href{http://dx.doi.org/10.1103/PhysRevB.98.134403}{Phys. Rev. B {\bfseries
  98}, 134403 (2018)}, [\href{https://arxiv.org/abs/1807.01322}{{\ttfamily
  arXiv:1807.01322}}].

\bibitem{Hislop:1981uh}
P.~D. Hislop and R.~Longo, \textit{{Modular Structure of the Local Algebras
  Associated With the Free Massless Scalar Field Theory}},
  \href{http://dx.doi.org/10.1007/BF01208372}{Commun. Math. Phys. {\bfseries
  84}, 71 (1982)}.

\bibitem{Casini:2011kv}
H.~Casini, M.~Huerta and R.~C. Myers, \textit{{Towards a derivation of  holographic entanglement entropy}},
  \href{http://dx.doi.org/10.1007/JHEP05(2011)036}{JHEP {\bfseries 05} (2011)
  036}, [\href{https://arxiv.org/abs/1102.0440}{{\ttfamily arXiv:1102.0440}}].

\bibitem{Wong:2013gua}
G.~Wong, I.~Klich, L.~A. Pando~Zayas and D.~Vaman, \textit{{Entanglement Temperature and Entanglement Entropy of Excited States}},
  \href{http://dx.doi.org/10.1007/JHEP12(2013)020}{JHEP {\bfseries 12} (2013)
  020}, [\href{https://arxiv.org/abs/1305.3291}{{\ttfamily arXiv:1305.3291}}].

\bibitem{Wen:2016inm}
X.~Wen, S.~Ryu and A.~W.~W. Ludwig, \textit{{Evolution operators in conformal  field theories and conformal mappings: Entanglement Hamiltonian, the
  sine-square deformation, and others}},
  \href{http://dx.doi.org/10.1103/PhysRevB.93.235119}{Phys. Rev. B {\bfseries
  93}, 235119 (2016)}, [\href{https://arxiv.org/abs/1604.01085}{{\ttfamily
  arXiv:1604.01085}}].

\bibitem{Cardy:2016fqc}
J.~Cardy and E.~Tonni, \textit{{Entanglement hamiltonians in two-dimensional conformal field theory}},
  \href{http://dx.doi.org/10.1088/1742-5468/2016/12/123103}{J. Stat. Mech.
  (2016) 123103}, [\href{https://arxiv.org/abs/1608.01283}{{\ttfamily
  arXiv:1608.01283}}].

\bibitem{Najafi:2016kwb}
K.~Najafi and M.~A. Rajabpour, \textit{{Entanglement entropy after selective  measurements in quantum chains}},
  \href{http://dx.doi.org/10.1007/JHEP12(2016)124}{JHEP {\bfseries 12} (2016)
  124}, [\href{https://arxiv.org/abs/1608.04074}{{\ttfamily
  arXiv:1608.04074}}].

\bibitem{Fries:2019acy}
P.~Fries and I.~A. Reyes, \textit{{Entanglement and relative entropy of a chiral fermion on the torus}},
  \href{http://dx.doi.org/10.1103/PhysRevD.100.105015}{Phys. Rev. D {\bfseries
  100}, 105015 (2019)}, [\href{https://arxiv.org/abs/1906.02207}{{\ttfamily
  arXiv:1906.02207}}].

\bibitem{Fries:2019ozf}
P.~Fries and I.~A. Reyes, \textit{{Entanglement Spectrum of Chiral Fermions on  the Torus}},
\href{http://dx.doi.org/10.1103/PhysRevLett.123.211603}{Phys. Rev. Lett. {\bfseries 123}, 211603 (2019)},
  [\href{https://arxiv.org/abs/1905.05768}{{\ttfamily arXiv:1905.05768}}].

\bibitem{Tonni:2017jom}
E.~Tonni, J.~Rodr\'{i}guez-Laguna and G.~Sierra, \textit{{Entanglement hamiltonian and entanglement contour in inhomogeneous 1D critical systems}},
  \href{http://dx.doi.org/10.1088/1742-5468/aab67d}{J. Stat. Mech. (2018)
  043105}, [\href{https://arxiv.org/abs/1712.03557}{{\ttfamily
  arXiv:1712.03557}}].

\bibitem{DiGiulio:2019lpb}
G.~Di~Giulio, R.~Arias and E.~Tonni, \textit{{Entanglement hamiltonians in 1D  free lattice models after a global quantum quench}},
  \href{http://dx.doi.org/10.1088/1742-5468/ab4e8f}{J. Stat. Mech. (2019)
  123103}, [\href{https://arxiv.org/abs/1905.01144}{{\ttfamily
  arXiv:1905.01144}}].

\bibitem{Kim:2016cdh}
P.~Kim, H.~Katsura, N.~Trivedi and J.~H. Han, \textit{{Entanglement and corner
  Hamiltonian spectra of integrable open spin chains}},
  \href{http://dx.doi.org/10.1103/PhysRevB.94.195110}{Phys. Rev. B {\bfseries
  94}, 195110 (2016)}, [\href{https://arxiv.org/abs/1512.08597}{{\ttfamily
  arXiv:1512.08597}}].

\bibitem{ParisenToldin:2018uzz}
F.~Parisen~Toldin and F.~F. Assaad, \textit{{Entanglement Hamiltonian of
  Interacting Fermionic Models}},
  \href{http://dx.doi.org/10.1103/PhysRevLett.121.200602}{Phys. Rev. Lett.
  {\bfseries 121}, 200602 (2018)},
  [\href{https://arxiv.org/abs/1804.03163}{{\ttfamily arXiv:1804.03163}}].

\bibitem{Kosior:2018vgx}
A.~Kosior, M.~Lewenstein and A.~Celi, \textit{{Unruh effect for interacting
  particles with ultracold atoms}},
  \href{http://dx.doi.org/10.21468/SciPostPhys.5.6.061}{SciPost Phys.
  {\bfseries 5}, 061 (2018)},
  [\href{https://arxiv.org/abs/1804.11323}{{\ttfamily arXiv:1804.11323}}].

\bibitem{Zhu:2019tsb}
W.~Zhu, Z.~Huang and Y.-C. He, \textit{{Reconstructing entanglement Hamiltonian
  via entanglement eigenstates}},
  \href{http://dx.doi.org/10.1103/PhysRevB.99.235109}{Phys. Rev. B {\bfseries
  99}, 235109 (2019)}, [\href{https://arxiv.org/abs/1806.08060}{{\ttfamily
  arXiv:1806.08060}}].

\bibitem{Turkeshi:2018hfx}
X.~Turkeshi, T.~Mendes-Santos, G.~Giudici and M.~Dalmonte,
  \textit{{Entanglement guided search for parent Hamiltonians}},
  \href{http://dx.doi.org/10.1103/PhysRevLett.122.150606}{Phys. Rev. Lett.
  {\bfseries 122}, 150606 (2019)},
  [\href{https://arxiv.org/abs/1807.06113}{{\ttfamily arXiv:1807.06113}}].

\bibitem{Mendes-Santos:2019ine}
T.~Mendes-Santos, G.~Giudici, R.~Fazio and M.~Dalmonte, \textit{{Measuring von
  Neumann entanglement entropies without wave functions}},
  \href{http://dx.doi.org/10.1088/1367-2630/ab6875}{New J. Phys. {\bfseries
  22}, 013044 (2020)}, [\href{https://arxiv.org/abs/1904.07782}{{\ttfamily
  arXiv:1904.07782}}].

\bibitem{Mendes-Santos:2019tmf}
T.~Mendes-Santos, G.~Giudici, M.~Dalmonte and M.~A. Rajabpour,
  \textit{{Entanglement Hamiltonian of quantum critical chains and conformal
  field theories}}, \href{http://dx.doi.org/10.1103/PhysRevB.100.155122}{Phys.
  Rev. B {\bfseries 100}, 155122 (2019)},
  [\href{https://arxiv.org/abs/1906.00471}{{\ttfamily arXiv:1906.00471}}].

\bibitem{fagotti2013reduced}
M.~Fagotti and F.~H. Essler, \textit{Reduced density matrix after a quantum
  quench}, \href{http://dx.doi.org/10.1103/PhysRevB.87.245107}{Phys. Rev. B
  {\bfseries 87}, 245107 (2013)},
  [\href{https://arxiv.org/abs/1302.6944}{{\ttfamily arXiv:1302.6944}}].

\bibitem{nielsen2010quantum}
M.~A. Nielsen and I.~L. Chuang, \textit{{Quantum Computation and Quantum
  Information}}.
\newblock Cambridge University Press, Cambridge, UK, 10th anniversary~ed.,
  2010,
  \href{http://dx.doi.org/10.1017/CBO9780511976667}{10.1017/CBO9780511976667}.

\bibitem{watrous2018theory}
J.~Watrous, \textit{{The Theory of Quantum Information}}.
\newblock Cambridge University Press, Cambridge, UK, 2018,
  \href{http://dx.doi.org/10.1017/9781316848142}{10.1017/9781316848142}.

\bibitem{Calabrese:2005in}
P.~Calabrese and J.~L. Cardy, \textit{{Evolution of entanglement entropy in
  one-dimensional systems}},
  \href{http://dx.doi.org/10.1088/1742-5468/2005/04/P04010}{J. Stat. Mech.
  (2005) P04010}, [\href{https://arxiv.org/abs/cond-mat/0503393}{{\ttfamily
  arXiv:cond-mat/0503393}}].

\bibitem{Calabrese:2006rx}
P.~Calabrese and J.~L. Cardy, \textit{{Time-dependence of Correlation Functions
  Following a Quantum Quench}},
  \href{http://dx.doi.org/10.1103/PhysRevLett.96.136801}{Phys. Rev. Lett.
  {\bfseries 96}, 136801 (2006)},
  [\href{https://arxiv.org/abs/cond-mat/0601225}{{\ttfamily
  arXiv:cond-mat/0601225}}].

\bibitem{Gilchrist2005distance}
A.~Gilchrist, N.~K. Langford and M.~A. Nielsen, \textit{{Distance measures to
  compare real and ideal quantum processes}},
  \href{http://dx.doi.org/10.1103/PhysRevA.71.062310}{Phys. Rev. A {\bfseries
  71}, 062310 (2005)},
  [\href{https://arxiv.org/abs/quant-ph/0408063}{{\ttfamily
  arXiv:quant-ph/0408063}}].

\bibitem{Liang:2018yey}
Y.-C. Liang, Y.-H. Yeh, P.~E. M.~F. Mendon\c{c}a, R.~Y. Teh, M.~D. Reid and
  P.~D. Drummond, \textit{{Quantum fidelity measures for mixed states}},
  \href{http://dx.doi.org/10.1088/1361-6633/ab1ca4}{Rept. Prog. Phys.
  {\bf 82}, 076001 (2019)},
  [\href{https://arxiv.org/abs/1810.08034}{{\ttfamily arXiv:1810.08034}}].

\bibitem{Zhang:2019wqo}
J.~Zhang, P.~Ruggiero and P.~Calabrese, \textit{{Subsystem Trace Distance in
  Quantum Field Theory}},
  \href{http://dx.doi.org/10.1103/PhysRevLett.122.141602}{Phys. Rev. Lett.
  {\bfseries 122}, 141602 (2019)},
  [\href{https://arxiv.org/abs/1901.10993}{{\ttfamily arXiv:1901.10993}}].

\bibitem{Zhang:2019itb}
J.~Zhang, P.~Ruggiero and P.~Calabrese, \textit{{Subsystem trace distance in
  low-lying states of $(1+1)$-dimensional conformal field theories}},
  \href{http://dx.doi.org/10.1007/JHEP10(2019)181}{JHEP {\bfseries 10} (2019)
  181}, [\href{https://arxiv.org/abs/1907.04332}{{\ttfamily
  arXiv:1907.04332}}].

\bibitem{Zhang:2019kwu}
J.~Zhang and P.~Calabrese, \textit{{Subsystem distance after a local operator
  quench}}, \href{http://dx.doi.org/10.1007/JHEP02(2020)056}{JHEP {\bfseries
  02} (2020) 056}, [\href{https://arxiv.org/abs/1911.04797}{{\ttfamily
  arXiv:1911.04797}}].

\bibitem{chehade2019quantum}
S.~S. Chehade and A.~Vershynina, \textit{Quantum entropies},
  \href{http://dx.doi.org/10.4249/scholarpedia.53131}{Scholarpedia {\bfseries
  14}, 53131 (2019)}.

\bibitem{Fannes1973}
M.~Fannes, \textit{A continuity property of the entropy density for spin
  lattice systems}, \href{http://dx.doi.org/10.1007/BF01646490}{Commun. Math. Phys. {\bfseries 31}, 291 (1973)}.

\bibitem{Audenaert:2006}
K.~M.~R. Audenaert, \textit{{A sharp continuity estimate for the von Neumann
  entropy}}, \href{http://dx.doi.org/10.1088/1751-8113/40/28/S18}{J. Phys. A {\bfseries 40}, 8127 (2007)},
  [\href{https://arxiv.org/abs/quant-ph/0610146}{{\ttfamily
  arXiv:quant-ph/0610146}}].

\bibitem{chen2017sharp}
Z.~Chen, Z.~Ma, I.~Nikoufar and S.~Fei, \textit{Sharp continuity bounds for
  entropy and conditional entropy},
  \href{http://dx.doi.org/10.1007/s11433-016-0367-x}{Sci. China Phys. Mech.
  Astron. {\bfseries 60}, 020321 (2017)},
  [\href{https://arxiv.org/abs/1701.02398}{{\ttfamily arXiv:1701.02398}}].

\bibitem{Holzhey:1994we}
C.~Holzhey, F.~Larsen and F.~Wilczek, \textit{{Geometric and renormalized
  entropy in conformal field theory}},
  \href{http://dx.doi.org/10.1016/0550-3213(94)90402-2}{Nucl. Phys. B
  {\bfseries 424}, 443 (1994)},
  [\href{https://arxiv.org/abs/hep-th/9403108}{{\ttfamily
  arXiv:hep-th/9403108}}].

\bibitem{Calabrese:2004eu}
P.~Calabrese and J.~L. Cardy, \textit{{Entanglement entropy and quantum field
  theory}}, \href{http://dx.doi.org/10.1088/1742-5468/2004/06/P06002}{J. Stat.
  Mech. (2004) P06002}, [\href{https://arxiv.org/abs/hep-th/0405152}{{\ttfamily
  arXiv:hep-th/0405152}}].

\bibitem{raggio1995properties}
G.~A. Raggio, \textit{Properties of q-entropies},
  \href{http://dx.doi.org/10.1063/1.530920}{J. Math. Phys. {\bfseries 36}, 4785 (1995)}.

\bibitem{chung2001density}
M.-C. Chung and I.~Peschel, \textit{Density-matrix spectra of solvable
  fermionic systems}, \href{http://dx.doi.org/10.1103/PhysRevB.64.064412}{Phys.  Rev. B {\bfseries 64}, 064412 (2001)},
  [\href{https://arxiv.org/abs/cond-mat/0103301}{{\ttfamily
  arXiv:cond-mat/0103301}}].

\bibitem{cheong2004many}
S.-A. Cheong and C.~L. Henley, \textit{{Many-body density matrices for free
  fermions}}, \href{http://dx.doi.org/10.1103/PhysRevB.69.075111}{Phys. Rev. B
  {\bfseries 69}, 075111 (2004)},
  [\href{https://arxiv.org/abs/cond-mat/0206196}{{\ttfamily
  arXiv:cond-mat/0206196}}].

\bibitem{Vidal:2002rm}
G.~Vidal, J.~I. Latorre, E.~Rico and A.~Kitaev, \textit{{Entanglement in
  Quantum Critical Phenomena}},
  \href{http://dx.doi.org/10.1103/PhysRevLett.90.227902}{Phys. Rev. Lett.
  {\bfseries 90}, 227902 (2003)},
  [\href{https://arxiv.org/abs/quant-ph/0211074}{{\ttfamily
  arXiv:quant-ph/0211074}}].

\bibitem{peschel2003calculation}
I.~Peschel, \textit{{Calculation of reduced density matrices from correlation
  functions}}, \href{http://dx.doi.org/10.1088/0305-4470/36/14/101}{J. Phys. A {\bfseries 36}, L205 (2003)},
  [\href{https://arxiv.org/abs/cond-mat/0212631}{{\ttfamily
  arXiv:cond-mat/0212631}}].

\bibitem{Latorre:2003kg}
J.~I. Latorre, E.~Rico and G.~Vidal, \textit{{Ground state entanglement in
  quantum spin chains}}, \href{http://dx.doi.org/10.26421/QIC4.1}{Quant. Inf.  Comput. {\bfseries 4}, 48 (2004)},
  [\href{https://arxiv.org/abs/quant-ph/0304098}{{\ttfamily
  arXiv:quant-ph/0304098}}].

\bibitem{peschel2012special}
I.~Peschel, \textit{{Special review: Entanglement in solvable many-particle
  models}}, \href{http://dx.doi.org/10.1007/s13538-012-0074-1}{Braz. J. Phys.
  {\bfseries 42}, 267 (2012)},
  [\href{https://arxiv.org/abs/1109.0159}{{\ttfamily arXiv:1109.0159}}].


\bibitem{Korepin:1994ui}
V.~E. Korepin, A.~G. Izergin, F.~H.~L. Essler and D.~B. Uglov,
  \textit{{Correlation function of the spin 1/2 XXX antiferromagnet}},
  \href{http://dx.doi.org/10.1016/0375-9601(94)90074-4}{Phys. Lett. A
  {\bfseries 190}, 182 (1994)},
  [\href{https://arxiv.org/abs/cond-mat/9403066}{{\ttfamily
  arXiv:cond-mat/9403066}}].

\bibitem{Essler:1994se}
F.~H.~L. Essler, H.~Frahm, A.~G. Izergin and V.~E. Korepin,
  \textit{{Determinant representation for correlation functions of spin 1/2 XXX  and XXZ Heisenberg magnets}},
  \href{http://dx.doi.org/10.1007/BF02099470}{Commun. Math. Phys. {\bfseries
  174}, 191 (1995)},
  [\href{https://arxiv.org/abs/hep-th/9406133}{{\ttfamily
  arXiv:hep-th/9406133}}].

\bibitem{Essler:1995vp}
F.~H.~L. Essler, H.~Frahm, A.~R. Its and V.~E. Korepin,
  \textit{{Integrodifference equation for a correlation function of the spin 1/2 Heisenberg XXZ chain}},
  \href{http://dx.doi.org/10.1016/0550-3213(95)00263-R}{Nucl. Phys. B  {\bf 446}, 448 (1995)},
  [\href{https://arxiv.org/abs/cond-mat/9503142}{{\ttfamily  arXiv:cond-mat/9503142}}].

\bibitem{shiroishi2001emptiness}
M.~Shiroishi, M.~Takahashi and Y.~Nishiyama, \textit{{Emptiness formation
  probability for the one-dimensional isotropic XY model}},
  \href{http://dx.doi.org/10.1143/JPSJ.70.3535}{J. Phys. Soc. Jpn. {\bf  70}, 3535 (2001)}, [\href{https://arxiv.org/abs/cond-mat/0106062}{{\ttfamily
  arXiv:cond-mat/0106062}}].

  \bibitem{Kitanine2002a}
 N. Kitanine, J.~M. Maillet, N.~A. Slavnov, V. Terras, 
 \textit{Emptiness formation probability of the XXZ spin-{$\f12$} Heisenberg chain at {\ensuremath{\Delta}} = {$\f12$}},
         \href{http://dx.doi.org/10.1088/0305-4470/35/27/102}{J. Phys. A {\bfseries 25}, L385 (2002)}
         [\href{https://arxiv.org/abs/hep-th/0201134}{{\ttfamily
  arXiv:hep-th/0201134}}].

\bibitem{Korepin2003}
V.~E. Korepin, S. Lukyanov, Y. Nishiyama, and M. Shiroishi, \textit{Asymptotic behavior of the emptiness formation probability in the critical phase of /XXZ spin chain},
\href{http://dx.doi.org/10.1016/S0375-9601(03)00616-9}{Phys. Lett. A {\bfseries 312}, 21 (2003)},
  [\href{https://arxiv.org/abs/cond-mat/0210140}{{\ttfamily
  arXiv:cond-mat/0210140}}].

\bibitem{Franchini:2005uv}
F.~Franchini and A.~G. Abanov, \textit{{Asymptotics of Toeplitz determinants
  and the emptiness formation probability for the XY spin chain}},
  \href{http://dx.doi.org/10.1088/0305-4470/38/23/002}{J. Phys. A  {\bfseries 38}, 5069 (2005)},
  [\href{https://arxiv.org/abs/cond-mat/0502015}{{\ttfamily  arXiv:cond-mat/0502015}}].

\bibitem{Stephan2013}
J-M St{\'e}phan, \textit{Emptiness formation probability, Toeplitz determinants, and conformal field theory},
\href{http://dx.doi.org/10.1088/1742-5468/2014/05/P05010}{J. Stat.  Mech. (2013) 05010}, [\href{https://arxiv.org/abs/1303.5499}{{\ttfamily
  arXiv:1303.5499}}].

\bibitem{najafi2016formation}
K.~Najafi and M.~A. Rajabpour, \textit{{Formation probabilities and Shannon
  information and their time evolution after quantum quench in the  transverse-field XY chain}},
  \href{http://dx.doi.org/10.1103/PhysRevB.93.125139}{Phys. Rev. B {\bfseries
  93}, 125139 (2016)}, [\href{https://arxiv.org/abs/1511.06401}{{\ttfamily  arXiv:1511.06401}}].

 \bibitem{Rajabpour2015}
 M. A. Rajabpour, \textit{Formation probabilities in quantum critical chains and Casimir effect},
 \href{http://dx.doi.org/10.1209/0295-5075/112/66001}{EPL {\bf 122}, 66001 (2015)}, [\href{https://arxiv.org/abs/1512.01052}{{\ttfamily
  arXiv:1512.01052}}].

 \bibitem{Viti2016}
 N. Allegra, J. Dubail, J-M. St\'ephan, and J. Viti, \textit{Inhomogeneous field theory inside the arctic circle}
 \href{http://dx.doi.org/10.1088/1742-5468/2016/05/053108}{J. Stat. Mech. (2016) 053108}, [\href{https://arxiv.org/abs/1512.02872}{{\ttfamily
  arXiv:1512.02872}}].

 \bibitem{Rajabpour2016}
M. A. Rajabpour, \textit{Finite size corrections to scaling of the formation probabilities and the Casimir effect in the conformal field theories},
\href{http://dx.doi.org/10.1088/1742-5468/2016/12/123101}{J. Stat. Mech. (2016) 123101}, [\href{https://arxiv.org/abs/1607.07016}{{\ttfamily
  arXiv:1607.07016}}].

 \bibitem{Ares2020}
F. Ares and J. Viti, \textit{Emptiness formation probability and Painlev\'e V equation in the XY spin chain}, 
\href{https://doi.org/10.10882F1742-54682Fab5d0b}{J. Stat. Mech. (2020) 013105}, [\href{https://arxiv.org/abs/1909.01270}{{\ttfamily
  arXiv:1909.01270}}].

\bibitem{Najafi:2019ypm}
M.~N. Najafi and M.~A. Rajabpour, \textit{{Formation probabilities and
  statistics of observables as defect problems in the free fermions and the
  quantum spin chains}}, 
\href{https://doi.org/10.1103/PhysRevB.101.165415}{PRB {\bf 101}, 165415 (2020)}  
  [\href{https://arxiv.org/abs/1911.04595}{{\ttfamily
  arXiv:1911.04595}}].

\bibitem{Slepian1978Prolate}
D.~{Slepian}, \textit{{Prolate spheroidal wave functions, Fourier analysis, and
  uncertainty. V - The discrete case}},
  \href{http://dx.doi.org/10.1002/j.1538-7305.1978.tb02104.x}{Bell Syst. Techn.
  J. {\bfseries 57}, 1371--1430 (1978)}.


\bibitem{Fagotti:2010yr}
M.~Fagotti and P.~Calabrese, \textit{{Entanglement entropy of two disjoint
  blocks in XY chains}},
  \href{http://dx.doi.org/10.1088/1742-5468/2010/04/P04016}{J. Stat. Mech.
  (2010) P04016}, [\href{https://arxiv.org/abs/1003.1110}{{\ttfamily
  arXiv:1003.1110}}].

\end{thebibliography}
\end{document}